\documentclass[fleqn,usenatbib]{mnras}

\usepackage{newtxtext,newtxmath}

\usepackage[T1]{fontenc}

\DeclareRobustCommand{\VAN}[3]{#2}
\let\VANthebibliography\thebibliography
\def\thebibliography{\DeclareRobustCommand{\VAN}[3]{##3}\VANthebibliography}

\usepackage{graphicx}	
\usepackage{amsmath}	

\newcommand{\gizmo}{\textsc{gizmo}}
\newcommand{\slug}{\textsc{slug}}
\defcitealias{Zhang2025}{Paper I}
\defcitealias{KT18}{KT18}

\newcommand{\aref}[1]{\hyperref[#1]{Appendix~\ref{#1}}}

\usepackage{scalerel,tikz}
\usetikzlibrary{svg.path}
\definecolor{orcidlogocol}{HTML}{A6CE39}
\tikzset{orcidlogo/.pic={
\fill[orcidlogocol] svg{M256,128c0,70.7-57.3,128-128,128C57.3,256,0,198.7,0,128C0,57.3,57.3,0,128,0C198.7,0,256,57.3,256,128z};
\fill[white] svg{M86.3,186.2H70.9V79.1h15.4v48.4V186.2z}
svg{M108.9,79.1h41.6c39.6,0,57,28.3,57,53.6c0,27.5-21.5,53.6-56.8,53.6h-41.8V79.1z M124.3,172.4h24.5c34.9,0,42.9-26.5,42.9-39.7c0-21.5-13.7-39.7-43.7-39.7h-23.7V172.4z}
svg{M88.7,56.8c0,5.5-4.5,10.1-10.1,10.1c-5.6,0-10.1-4.6-10.1-10.1c0-5.6,4.5-10.1,10.1-10.1C84.2,46.7,88.7,51.3,88.7,56.8z};
}}
\newcommand\orcidicon[1]{\href{https://orcid.org/#1}{\mbox{\scalerel*{
\begin{tikzpicture}[yscale=-1,transform shape]
\pic{orcidlogo};
\end{tikzpicture}
}{|}}}}

\title[Ripples of Stellar Enrichment]{Ripples of Stellar Enrichment (RoSE) -- simulating element production and mixing in a Milky Way-mass galactic disc star-by-star}

\author[C. Zhang et al.]{\ignorespaces
Chuhan Zhang$^{\orcidicon{0009-0000-2503-4803}{1}}$\thanks{E-mail: chuhan.zhang@anu.edu.au},
Mark R. Krumholz$^{\orcidicon{0000-0003-3893-854X}{1}}$,
Melissa K. Ness$^{\orcidicon{0000-0001-5082-6693}{1}}$,
Yuan-Sen Ting$^{\orcidicon{0000-0001-5082-9536}{2,3}}$,
Zefeng Li$^{\orcidicon{0000-0001-7373-3115}{4}}$,
and
\newauthor
Zipeng Hu$^{\orcidicon{0000-0002-3758-552X}{5}}$
\\
$^{1}$Research School of Astronomy and Astrophysics, Australian National University, Cotter Road, Weston ACT 2611, Australia\\
$^{2}$Department of Astronomy, The Ohio State University, Columbus, OH 43210, USA\\
$^{3}$Max-Planck-Institut f\"ur Astronomie, K\"onigstuhl 17, D-69117 Heidelberg, Germany\\
$^{4}$Centre for Extragalactic Astronomy, Department of Physics, Durham University, South Road, Durham DH1 3LE, UK\\
$^{5}$Kavli Institute for Astronomy and Astrophysics, Peking University, 5 Yiheyuan Road, Haidian District, Beijing 100871, People’s Republic of China
}

\date{Accepted XXX. Received YYY; in original form ZZZ}

\pubyear{\the\year{}}

\begin{document}
\label{firstpage}
\pagerange{\pageref{firstpage}--\pageref{lastpage}}
\maketitle

\begin{abstract}
We present the Ripples of Stellar Enrichment (RoSE) simulations, which follow an isolated Milky Way-mass disc galaxy with star-by-star feedback and nucleosynthesis from five physical enrichment channels -- Wolf-Rayet stars, type II supernovae, type Ia supernovae, asymptotic giant branch stars, and neutron star mergers. We use these simulations to test how elements' diverse nucleosynthetic origins imprint spatial, temporal, and inter-element abundance correlations in gas and newly formed stars. We find that nucleosynthetic source composition is the primary organising principle of elemental structure: elements sharing a dominant production channel exhibit similar spatial and temporal statistics and their abundances are strongly correlated with one another, while mixed-source pairs are much more weakly correlated. We show that a simple linear regression model based only on how element pairs differ in their nucleosynthetic origin is able to predict, with high fidelity, how strongly their abundances correlate, in both interstellar medium gas and coeval stars. Comparison with Milky Way stellar abundance data shows encouraging qualitative agreement, with differences between simulations and observations comparable to the scatter between independent observational datasets. Together, these results show that the covariance of galactic abundances is largely predictable from the mixture of nucleosynthetic sources supplying each element: same-source elements remain tightly correlated in gas and young stars, whereas differing source mixtures produce systematically weaker correlations.
\end{abstract}




\section{Introduction}
 
The distribution of metals (elements heavier than helium) within galaxies provides a critical window into the physics of the interstellar medium (ISM) and galaxy assembly. Metals are produced primarily through stellar nucleosynthesis, with stars of different masses producing different elements over different timescales, and then injected into the ISM through stellar winds, supernovae (SNe), and compact object mergers. Once in the ISM, these metals are mixed by turbulence, transported by large-scale flows, and incorporated into new generations of stars, driving the chemical evolution of galaxies and shaping the conditions for future star formation (for reviews, see \citealt{Tinsley_1980, Maiolino_2019, Sanchez_2021}).

The past two decades have witnessed a revolution in our ability to map metal distributions spatially. The metallicity of the ISM, most commonly traced via oxygen abundance in ionised regions, can be measured using emission line diagnostics (for a review, see \citealt{Kewley_2019}). The deployment of integral field units (IFUs) enables measurements of spatially resolved two-dimensional distributions of oxygen abundance across nearby galaxies (e.g., \citealt{2011MNRAS.415.2439R, 2016ApJ...830L..40S, Huang2026}). These observations reveal metallicity gradients, typically showing that metal abundance decreases from centres of galaxies outward (e.g., \citealt{2017MNRAS.469..151B, 2018A&A...618A..64H, 2018MNRAS.479.5235P}). A range of theoretical studies have aimed to explain these gradients and situate them in the broader context of galaxy formation (e.g., \citealt{2009A&A...499..427D, 2017MNRAS.466.4780M, 2021MNRAS.502.5935S}).

However, gradients represent a significant simplification of the data, since they collapse complex two-dimensional (2D) maps down to a single inside-out linear fit. To exploit the full information content of IFU metallicity maps, higher-order statistics are needed to decode the detailed processes of metal injection and mixing. One of the simplest yet most informative statistics for 2D maps is the two-point correlation function, which quantifies characteristic length scales over which metallicity fluctuations persist. \citet[hereafter \citetalias{KT18}]{KT18} provide a minimal theoretical framework to predict two-point correlations of galaxy metallicities based on the competition between metal injection and diffusion. This prediction has motivated various observational studies examining two-point correlations (or similar statistics) in nearby galaxies (e.g., \citealt{2020MNRAS.499..193K, Li2021, 2023MNRAS.518..286L, 2021MNRAS.508..489M, 2022MNRAS.509.1303W, 2024arXiv240704252L}). These studies find that metallicity maps of nearby galaxies contain statistically significant spatial structure beyond the overall gradient, and that the two-point correlation functions describing this structure generally follow the shape predicted by \citetalias{KT18}, with correlation lengths of $\sim 1$ kpc, though with significant systematic variations with galaxy properties such as stellar mass and star formation rate.

Very recently, two-dimensional metallicity maps have also started to become available for other elements, most notably nitrogen and sulfur \citep{Fabio25, Li2025}, enabling both comparisons of the spatial statistics of different elements and calculations of the extent of element-element cross-correlation. \citet{KT25} extended the original \citetalias{KT18} theory to incorporate such statistics. The models are qualitatively consistent with the data thus far, but to date there are only two galaxies with measured two-dimensional abundance maps for multiple elements. This situation is likely to improve as facilities such BlueMUSE come online \citep{BlueMUSE}, providing access to key diagnostic lines for nitrogen and potentially other elements.

Stars within the Milky Way provide another, complementary, observational window on elemental abundance distributions. Only a small number of elements, all produced primarily by either core collapse SNe or AGB stars, can be measured in the gas. By contrast, stellar spectra provide access to a much larger suite of elements. 
While such measurements do not directly trace the spatial distribution of elements in galaxies, since stars migrate after formation, any spatial correlations between elements that were present at formation remain frozen into the distribution of stars in chemical abundance space (excluding fully convective stars whose surface abundances can be altered by internal processes). There have been extensive efforts to understand this chemical space structure, since it matters for a wide variety of studies relying on stellar abundances \citep{Bland-Hawthorn10a, Bland-Hawthorn16a, Krumholz19a, Weinberg22a, Ting22a, Manea2025, Mead25a}, and the data are again qualitatively consistent with the analytic \citet{KT25} model.

While analytic and observational work has begun the study of metallicity distribution statistics, simulation efforts have been limited. \citet{2024MNRAS.528.7103L} post-processed the Auriga cosmological simulations \citep{2017MNRAS.467..179G} to produce metallicity maps comparable to those accessible via observations, finding that simulations successfully reproduce observed correlation lengths and are in reasonable agreement with \citetalias{KT18} predictions. In \citet[hereafter \citetalias{Zhang2025}]{Zhang2025} we extended this work using isolated galaxy simulations with mass resolution more than an order of magnitude better than Auriga, and star-by-star tracking of feedback and nucleosynthesis, enabling us to follow metal return by individual stellar sources. This additional resolution allows us to model potentially important mixing associated with small-scale phenomena such as spiral arms \citep{2016MNRAS.460L..94G, 2023MNRAS.521.3708O} and thermal instability \citep{Yang_2012} more explicitly, complementing cosmological simulations that capture galaxy assembly and environmental effects. \citetalias{Zhang2025} demonstrated that small-scale metallicity fluctuations naturally organize into nucleosynthetic families, with elements from the same production channel (e.g., O, Mg, S from core-collapse supernovae) correlating more strongly with each other than with elements from different channels, consistent with the results of observations of both gas and stars \citep{Ting22a, Li2025, Mead25a}.

However, the work in \citetalias{Zhang2025} left significant gaps. First, due to its high resolution, the simulation used in that paper could only be run for $\sim 0.5$ Gyr, and as a result the paper omitted elements produced via nucleosynthetic channels that require longer timescales to return their products, most prominently iron peak and $s$- and $r$-process nuclides. Second, the simulation code used in that paper only recorded elemental abundances in gas instead of newly-produced stars limiting its utility for making comparisons with the growing body of stellar chemical abundance space measurements. 

These gaps motivate the present study, which extends the high-resolution isolated Milky Way-mass disc simulation framework established in \citetalias{Zhang2025} to address multiple outstanding questions simultaneously. Building on our star-by-star treatment of feedback and nucleosynthesis, we now (1) expand coverage to six modelled nucleosynthetic channels including Type Ia supernovae and neutron star mergers, enabling complete tracking of nine isotopes spanning short and long delay times; (2) extend the simulation duration to 600 Myr, which is long enough for the correlation statistics we analyse to settle into a quasi-steady state; (3) implement component-resolved correlation analysis to disentangle contributions from different channels; (4) incorporate cross-correlation analysis between element pairs to identify whether differences in the spatial distributions of particular elements are driven by their fractional contributions from different production channels; and (5) track chemical inheritance into newly formed stars to connect gas-phase and stellar abundance patterns. We use the resulting data to test the extended \citet{KT25} framework against high-resolution simulations, examine how nucleosynthetic diversity manifests in both spatial and temporal correlation statistics, and explore the physical mechanisms linking chemical injection, gas cycling between ISM phases, and spatial mixing in the ISM. In doing so, we provide first-principles predictions for the structure of both gas-phase and stellar chemical abundance space that can be compared to current and future IFU surveys and stellar spectroscopic campaigns.

\section{Simulation}

We present the second generation of our numerical simulations, Ripples of Stellar Enrichment (RoSE), developed from the isolated, magnetised, hydrodynamic $L_\ast$ disc galactic chemical simulation presented in \citetalias{Zhang2025}. This new simulation retains the star-by-star tracking method for chemical feedback introduced in our previous work but includes a broader range of elements and nucleosynthetic sources. In this section, we only summarise most aspects of the simulations, and go into detail only when discussing specific improvements and key differences between the two generations of simulations. For full details regarding the numerical setup and methodology of the first-generation simulations, we refer the reader to \citetalias{Zhang2025}.

\subsection{Initial conditions}
\label{ssec:initcon}

As in the first-generation simulations, we begin from the 600 Myr snapshot from \cite{WK23}; these authors used the \gizmo~code \citep{gizmo} to simulate an isolated galaxy with initial conditions taken from the AGORA project \citep{Kim16b}, which was designed to be a close match to the bulk properties of the Milky Way. The simulation includes a gaseous disc, a stellar disc, a stellar bulge, and a dark matter halo, all of which are simulated live rather than using fixed potentials. However, we caution that, while the simulation matches gross Milky Way properties such as stellar mass, gas mass, and bulge-to-disc ratio, we do not attempt to reproduce more detailed aspects of the Milky Way's structure such as the properties of the bar, interactions with satellite galaxies, or the effects of external accretion; we defer a full discussion of these caveats and their implications to \autoref{ssec:caveats}.

\cite{WK23} used the native implementation of stellar feedback in \gizmo~\citep{gizmo}, which was appropriate for their $\approx 10^3$ M$_\odot$ resolution, but is insufficient for the much higher resolution we seek to reach. Our first step therefore is to restart the simulation replacing the \gizmo~feedback module with one that uses star-by-star-tracking based on the \slug~\citep[Stochastically Lighting Up Galaxies;][]{slug1, slug2} stellar population synthesis code. In this approach for each star particle formed we draw a stellar population for that particle star-by-star from the initial mass function, and then inject ionising and supernova feedback based on star-by-star calculations of ionising luminosity and timing of supernova explosions. We use this method to continue the simulation for an additional 100 Myr, allowing it to re-settle into statistical steady-state with the new feedback approach.

The second step is that, following the approach outlined in \cite{Hu23}, we increase the mass resolution to 286.4 M$_\odot$ and reduce the minimum gas and particle softening lengths to their final values (see next section), then evolve the system for a further 2 Myr, long enough for the system to settle back to statistical steady-state at the increased resolution. The final state at the conclusion of this relaxation period serves as the initial condition for the RoSE project. In the discussion that follows, we will take this state to define $t=0$, and we run the simulation for an additional 600 Myr, with outputs saved at a cadence of 1 Myr.

\subsection{Numerical methods}
\label{ssec:methods}

\subsubsection{Simulation physics: MHD, cooling, star formation, and feedback}

We run our simulation using \gizmo's meshless finite-mass (MFM) Godunov solver with ideal MHD enabled and constrained-gradient divergence control \citep{gizmo, Hopkins16b, Hopkins16a}. The simulation includes multiple metal fields (see below for details), each of which is treated as a passive scalar. In addition to advective transport with the gas, these scalars are diffused using \gizmo's default Smagorinsky turbulent model \citep{Colbrook17a, Hopkins18b}. The simulation includes a simple non-equilibrium chemical network to follow H and He ionisation, and uses the \textsc{Grackle} module \citep{Smith17a} to handle radiative cooling and heating; we use this rather than the default \gizmo~cooling module because the latter does not properly produce a two-phase neutral medium (see \citealt{WK23} for details). Gravity and hydrodynamics use adaptive softening with a minimum 0.0001 code units for gas ($\approx0.1$ pc physical), 0.005 for ``active'' star particles formed during the simulation, 0.05 for dark matter, and 0.02 for ``passive'' star particles present from the start of the simulation.

Star formation in the simulation begins once gas reaches a density threshold $\rho_\mathrm{c}=10^3$ H atoms per cm$^3$, which at our final resolution corresponds to a density for which the particle mass is approximately equal to the Jeans mass for gas in thermal equilibrium. For gas above this threshold, the per-freefall star formation efficiency is $\epsilon_\mathrm{ff}=0.01$, consistent with observations \citep[and references therein]{Krumholz19a}. To conserve computational resources, we increase this to 100 (i.e., immediate conversion to stars guaranteed) for gas exceeding 100 times the density threshold. As mentioned above, when a star particle forms, we use \slug~\citep{slug1, slug2} to draw a population of individual stars that will populate that star particle, and whose evolution will determine its feedback thereafter. We use the same IMF, tracks, and stellar atmospheres in \slug~as in \citetalias{Zhang2025}. Stars feed back on the gas via photoionisation and supernovae. The photoionising luminosities of each star particle are computed on-the-fly by \slug, and we implement photoionisation using a Str\"omgren volume approximation. Similarly, \slug~identifies which individual stars explode as supernovae using the models of \citet{Sukhbold2016}, and provides the time after formation at which these explosions occur; each explosion adds $10^{51}$ erg to the gas, which we implement using a mixed momentum-energy feedback model since even at our high resolution the Sedov-Taylor phase is often unresolved. See \citet{Armillotta2019} and \citet{WK23} for details on how feedback is implemented.

\subsubsection{Metal injection and tracing}

The RoSE project introduces three major functional enhancements over its predecessor. The first is that, rather than simply tracking the total abundance of each isotope we follow, we also decompose the contributions from each modelled nucleosynthetic source, allowing us to identify how much each nucleosynthetic channel has contributed to the abundance of each isotope at any given position. The sources we model that are the same as in \citetalias{Zhang2025} are Wolf-Rayet stellar winds (WR), Type II supernovae (SNII), and asymptotic giant branch stellar winds (AGB). The second is that we include two additional nucleosynthetic channels: Type Ia supernovae (SNIa) and neutron star mergers (NSM). As part of the implementation of these additional sources, we have also improved our treatment of AGB yields. The third major improvement is the implementation of chemical inheritance, where newly formed stars inherit the chemical abundances of the gas from which they form. This provides a self-consistent framework within which we can simultaneously study the chemical evolution of the stellar and gaseous components within the galaxy. In the remainder of this section we detail the implementation of each of these improvements.

Within the RoSE framework, we independently track nine isotopes -- which we list in \autoref{tab:old_star_yields} -- produced by six distinct nucleosynthetic channels. We track each combination of element and channel as a distinct scalar field, so we in the analysis that follows we can reconstruct not just the abundances of all elements, but the relative contributions to those abundances made by different channels. Given that our simulation runs for $\lesssim 1$ Gyr, we can broadly divide these channels into two categories: those that return elements on timescales comparable to or smaller than the duration of our simulation, and where therefore all of the element return is due to stars that form self-consistently during the simulation, and those that operate on longer timescales and where element return must therefore be attributed to stars that were present in the simulation initial condition. We use distinct methods to treat these two channels. In the discussion that follows and in the remainder of this paper, we will use the term ``\slug~particles'' to refer to star particles that form self-consistently over the course of the simulation, which have \slug~stellar populations associated with them, and ``old star particles'' to refer to star particles that were present prior to $t=0$, and which do not have associated \slug~stellar populations.

\paragraph{Short-timescale nucleosynthetic channels.}
Our short-timescale nucleosynthetic channels include enrichment from WR stars, SNII explosions, and massive AGB stars. For these channels element return comes from \slug~particles, and our numerical implementation is as follows. Each \slug~particle has an associated stellar population drawn from the IMF, and each star drawn in turn has an individual nucleosynthetic yield, which is taken from tabulations of the yields for particular channels. We take our yields for WR winds and SNII explosions from \citet{Sukhbold2016}, while we take our yields for AGB stars from \citet{Doherty2014} and \citet{Karakas2016}; see \citetalias{Zhang2025} for details. Note that only some AGB stars return yields on timescales shorter than our simulation duration; we discuss how we divide the AGB population based on its lifetime below.

For WR winds and SNII, element return occurs over a short timescale relative to the dynamical time of the simulation, and we therefore retain the scheme used in \citetalias{Zhang2025}: when a star reaches the end of its life, we instantaneously add the elements in returns to the surrounding gas particles using the same weighting kernel as that used to distribute supernovae energy and momentum. In contrast, AGB stars contribute feedback through a continuous mass loss process over extended timescales. Implementing this in a truly continuous manner would be computationally prohibitive, so we use an approximate method wherein the AGB yield is calculated and injected into the ISM only at the moment its parent \slug~particle is dynamically awakened by the simulation code. During this awakening phase, the code checks the particle's evolutionary history and computes the total integrated AGB yield accumulated since the last awakening event. These awakenings happen on a locally-determined dynamical timescale, ensuring that gas deposition happens on the same timescale as gas or stellar motion.

\paragraph{Long-timescale nucleosynthetic channels.}
The long-timescale return channels are SNe Ia, NSMs, and low-mass AGB stars, and we assign mass return via these channels to old star particles. For SNe Ia and NSMs, we implement a stochastic model whereby we assign a specific occurrence rate (i.e., rate per unit mass) for each event, which we apply probabilistically. For SNIa, the observed rate in Milky Way-mass galaxies is approximately 0.5 event per century \citep{Li2011b, Wiseman2021}, and our simulation initial condition has a stellar mass of $3.7\times 10^{10}$ M$_\odot$, so we set the rate per unit time per unit stellar mass to $\Gamma_\mathrm{Ia} = 1.4\times 10^{-13}$ M$_\odot^{-1}$ yr$^{-1}$. Thus during each update through time $\Delta t$ for an old star particle of mass $M_*$, we assign it a probability $P = 1 - \exp(-\Gamma_\mathrm{Ia} M_* \Delta t)$ of hosting a SNIa.

We handle NSMs in exactly the same way, except that the rate of NSMs is considerably lower. Although the most commonly cited estimate for the Milky Way is on the order of 10 events per million years, corresponding to 20\% of stars with $8-20$ $\text{M}_\odot$ eventually producing NSMs \citep{Hirai2022}, we enhance this rate to 500 events per million years in our model, corresponding to $\Gamma_\mathrm{NSM} = 1.4\times 10^{-14}$ M$_\odot^{-1}$ yr$^{-1}$ to ensure that a statistically significant sample of NSM events occur during the simulation timeframe, allowing for a robust investigation of their distinct contribution to the galactic chemical inventory, particularly for elements like r-process nuclei; we reduce the yield per event correspondingly, so that the total element production rate remains the same.

For both NSM and SNIa, once an explosion event has been triggered in an old stellar particle, we then inject $10^{51}$ erg of energy, along with $1.4$ $\text{M}_\odot$ of mass for SNIa and $0.18 \text{M}_\odot$ for NSM, to the surrounding ISM. For the SNIa channel, we adopt element-by-element yields per event $\Delta M_i$ from \citet{Seitenzahl2013}, specifically their Table 2 for the case of a progenitor metallicity $Z = 0.02$. For the NSM channel, we use the yield calculations from \cite{Holmbeck2024}, again diluting down by the same factor by which we have enhanced the rates so that the total yield is the same. We summarise the mass return per event for each element for each channel in \autoref{tab:old_star_yields}. We distribute the injected mass from these channels to fluid elements surrounding the triggering particle exactly as we do for SNII.

For old AGB stars, which inject mass near-continuously, we inject mass from each old star particle each time it is awakened during the simulation evolution, using the same strategy as for AGB yields from \slug~particles. To determine the rate of element return, we use \slug~to calculate the cumulative yield $\Delta M_i$ of each element $i$ from a simple stellar population of mass $M_\mathrm{SSP}$ over a time $t_\mathrm{H}=13.6$ Gyr, using the same IMF, tracks, and yield tables as for \slug~particles, and disabling all nucleosynthetic mechanisms except AGB stars. We then set the element return rate per unit mass for old star particles to $\dot{y}_i = \Delta M_i / M_\mathrm{SSP} t_\mathrm{H}$, which is equivalent to assigning each old star particle a mass return rate equal to the time-averaged return rate over a Hubble time. This is a crude approximation, since it ignores the detailed star-formation history of the galaxy, so in \aref{app:oldagb_sensitivity} we quantify its impact by repeating the elemental cross-correlation analysis after multiplying the old-star AGB contribution by factors $f=0$, $0.5$, $1$, and $2$. This post-processing test brackets both removal and a factor-of-two enhancement of the adopted rate.

We summarise our old star yields in \autoref{tab:old_star_yields}. For the long-timescale channels associated with old star particles, the element-yield vector for each channel is fixed by construction. We therefore track one scalar field per long-timescale channel, and reconstruct the isotope-by-isotope contribution by multiplying that scalar by the fixed yield ratios in \autoref{tab:old_star_yields}. By contrast, the short-timescale channels associated with newly formed \slug~particles are sampled stochastically from the IMF, so two star particles with the same total mass need not return elements in identical ratios. For those channels we therefore track separate scalar fields for each combination of isotope and nucleosynthetic channel.

\begin{table} 
    \centering
    \caption{Old star yield table used in RoSE. For SNIa and NSM we report the mass returned per event; SNIa values are taken from \citet{Seitenzahl2013}, and NSM values from \citet{Holmbeck2024}. For old AGB stars, the quantity listed is the mass return per unit stellar mass unit time.}
    \label{tab:old_star_yields}
    \renewcommand{\arraystretch}{1.4}
    \begin{tabular}{cccc} 
        \hline
        Isotope & SNIa [M$_\odot$] & NSM [M$_\odot$] & old AGB [Myr$^{-1}$] \\
        \hline
        $^{12}$C   & $3.04\times10^{-3}$ & 0 & $1.00\times10^{-7}$ \\
        $^{14}$N   & $3.21\times10^{-6}$ & 0 & $4.71\times10^{-8}$ \\
        $^{16}$O   & $1.01\times10^{-1}$ & 0 & $1.15\times10^{-7}$ \\
        $^{24}$Mg  & $1.52\times10^{-2}$ & 0 & $1.14\times10^{-8}$ \\
        $^{32}$S   & $1.11\times10^{-1}$ & 0 & $6.05\times10^{-9}$ \\
        $^{56}$Fe  & $6.22\times10^{-1}$ & 0 & $2.45\times10^{-8}$ \\
        $^{138}$Ba & 0 & $9.2\times10^{-7}$ & $3.30\times10^{-12}$ \\
        $^{140}$Ce & 0 & $2.5\times10^{-7}$ & $9.21\times10^{-13}$ \\
        $^{153}$Eu & 0 & $1.1\times10^{-7}$ & $6.27\times10^{-15}$ \\
        \hline
    \end{tabular}
\end{table}

\section{Analysis methods}

This paper extends the two-point correlation toolkit developed in \citet{Li2021} and \citetalias{Zhang2025} to (i) quantify spatial and temporal scales of mixing and (ii) diagnose how nucleosynthetic channels imprint distinct correlation signatures across elements. In this section we detail the steps in our analysis pipeline.

\subsection{Map-making and pre-processing}
\label{ssec:map_making}

Our raw outputs consist of 600 snapshots files, saved at a cadence of 1 Myr; each file contains fields storing the abundance of each element in each gas and \slug~particle. Our first processing step is that, for each snapshot and element, we construct a face-on 2D map over a 20 kpc $\times$ 20 kpc square centred on the galaxy centre\footnote{Note that in \citetalias{Zhang2025} we used a 30 kpc $\times$ 30 kpc region. However, for the present paper we found that this introduces some noise coming from regions further from the galactic centre where the particle distribution is more poorly sampled. We have therefore restricted our analysis region to be somewhat closer to the galactic centre.}, sampled on an 800 $\times$ 800 grid ($\Delta x = \Delta y = 25$ pc). For each pixel in the map we compute surface densities of total gas mass and mass of each element, $\Sigma$ and $\Sigma_Z$. We compute the elemental surface densities both for the total mass of each element listed in \autoref{tab:old_star_yields} summed over all nucleosynthetic sources, and for the individual fields broken down by source. For each element we define the log-metallicity field as
\begin{equation}
    m = \log \frac{\Sigma_Z}{\Sigma}.
\end{equation}
To remove the dominant radial trend, we subtract the azimuthal median profile in annuli of width $\Delta r=25$ pc to obtain the metal fluctuation field,
\begin{equation}
    m' = m - \overline{m}(r),
\end{equation}
where $\overline{m}(r)$ is the median metallicity in each radial bin. Finally, to avoid biases due to poorly-sampled regions at large radii, we mask pixels with $\Sigma < 1\,\mathrm{M}_\odot\,\mathrm{pc}^{-2}$. We refer to the final product of this procedure for each element $a$, $m'_a$, as the metal fluctuation map, and all the statistics we describe below are computed from the non-masked pixels of this map.

\subsection{Spatial autocorrelation}
\label{ssec:analysis_acorr}

To quantify spatial structure of each element, we next compute the autocorrelation for element $a$ as
\begin{equation}
    \xi_{a,a}(r) = \left\langle \frac{\langle m_a'(\mathbf{r'}+\mathbf{r})\,m_a'(\mathbf{r'}) \rangle_{\mathbf{r}'}}{\langle m_a'^2(\mathbf{r'}) \rangle_{\mathbf{r}'}} \right\rangle_{\theta},
    \label{eq:auto_corr}
\end{equation}
where the outer average over $\theta$ is the mean over all orientations of the position vector $\mathbf{r}$, and the inner averages over $\mathbf{r}'$ are averages over all positions. In practice we evaluate this quantity by considering every pixel pair in the map and binning pairs by separation. Let $r_n$ be the central separation for bin $n$ with edges $r_{n\pm 1/2}$. Then we compute the autocorrelation at separation $r_n$ as
\begin{equation}
    \xi_{a,a}(r_n) = \frac{1}{\sigma_a^2}\,\frac{1}{N_n}\sum_{r_{n-1/2} < r_{ij} < r_{n+1/2}} m_a'(\mathbf{r}_i)\,m_a'(\mathbf{r}_j),
\end{equation}
where $\sigma_a^2 = \langle m_a'^2\rangle$ is the variance over the total metal fluctuation field and $N_n$ is the number of pairs in the bin. We evaluate the autocorrelation over bins of width $\Delta r = 25$ pc from separations of zero (where $\xi_{a,a} = 1$ by construction) up to the maximum scale allowed by the 20 kpc map. We estimate uncertainties in the autocorrelation using a jackknife procedure of dividing the map into 12 equal-area azimuthal sectors; the jackknife variance across sectors provides conservative errors on $\xi_{a,a}(r)$.

In addition to the raw auto-correlation, we also compute parameters for a parametric fit to the autocorrelation data. In \citetalias{Zhang2025}, we showed that the autocorrelation data are reasonably well-fit by the analytic model proposed by \citetalias{KT18}, which predicts a functional form for the autocorrelation function
\begin{equation}
\xi(r) = \frac{2}{\ln \left( 1 + 2 \phi^2 \right)}
\int_0^\infty e^{-l_\mathrm{corr}^2 a^2/\phi^2}\left( 1 - e^{-2 l_\mathrm{corr}^2 a^2} \right)\frac{J_0(ar)}{a} \, da,
\label{eq:kt18-space}
\end{equation}
where $J_0$ is the zeroth-order Bessel function, $l_\mathrm{corr}$ is the correlation length describing the large-scale structure of the autocorrelation function, and $\phi = l_\mathrm{corr}/x_0$ is the ratio of the correlation length to $x_0$, the effective size of the region of which elements are initially deposited by stellar feedback. 

For each element, source, and snapshot we fit $x_0$ and $l_\mathrm{corr}$ by comparing our measured autocorrelations to the predictions of \autoref{eq:kt18-space} using the \textsc{emcee} MCMC sampler  \citep{Foreman-Mackey13a}. We adopt uniform priors $0<x_0/\mathrm{pc}<10^4$, $0<l_\mathrm{corr}/\mathrm{pc}<10^{5}$, and take our log likelihood function to be $\log\mathcal{L} = -\sum_n [\xi_{a,a}(r_n) - \xi_\mathrm{mod}(r_n)]^2 / 2 \sigma_n^2$, where $\xi_{a,a}(r_n)$ is the autocorrelation we measure at separation $r_n$ from the map, $\xi_\mathrm{mod}(r)$ is evaluated from \autoref{eq:kt18-space}, $\sigma_n$ is the variance for bin $n$ that we determine via our jackknife procedure, and the sum runs over all bins up to the first zero-crossing of $\xi_{a,a}(r_n)$. These parts of our fitting procedure are identical to that in \citetalias{Zhang2025}, so we refer readers to that paper for further details. To obtain our final posterior distributions, we run \textsc{emcee} using 100 walkers with 1000 steps, discarding the first 100 steps for burn-in and thinning the remaining parts of the chains by a factor of 5.

\subsection{Time-correlation}
\label{ssec:analysis_tcorr}

To measure how rapidly metal structures decorrelate in time, we compute a time correlation, compensated for the effects of differential rotation. To compute this quantity, we first tabulate a gas rotation curve $v_c(r)$ from the simulation by computing the mass-weighted circular velocity in 25 pc radial bins; we take the mean over all snapshots for this purpose, but the variance from one snapshot to another is very small. Then in order to compute the correlation between the metal fluctuation fields $m_a'(\mathbf{r},t)$ and $m_a'(\mathbf{r},t + \tau)$ separated by a time lag $\tau$, we shift the positions of all particles in the later snapshot to remove the effects of galactic rotation; in practice we accomplish this by converting the particles' projected positions $(x,y)$ in the galactic plane to a polar coordinate system $(r, \theta)$ with its origin at the galactic centre, changing the $\theta$ coordinate to $\theta' = \theta - v_c(r)\tau/r$, and then converting the resulting $(r,\theta')$ coordinates back to Cartesian positions. We then compute a de-rotated metal fluctuation field $\widetilde{m}_a'(\mathbf{r},t+\tau)$ from the updated positions, using the same procedure as described in \autoref{ssec:map_making}. 

We now define the (zero-lag-in-space) time correlation for element $a$ as
\begin{equation}
    \xi_a(\tau) = \frac{\langle m_a'(\mathbf{r},t)\, \widetilde{m}_a'(\mathbf{r},t+\tau) \rangle_\mathbf{r}}{\sigma_a(t)\,\sigma_a(t+\tau)},
    \label{eq:time_corr_zerolag}
\end{equation}
where $\sigma_a(t)$ is the total variance of the metal fluctuation map $m'_a$ at time $t$. In practice we compute this quantity from our discrete, pixelised maps as
\begin{equation}
    \xi_a(\tau) = \frac{\sum_i m'_{a,i}(t) \widetilde{m}'_{a,i}(t+\tau)}{\sqrt{\sum_i [m'_{a,i}(t) \widetilde{m}'_{a,i}(t+\tau)]^2}},
\end{equation}
where $m_{a,i}$ is the value the metal fluctuation map in pixel $i$, and the sums run over all non-masked pixels. While in principle we can compute this quantity for any pair of snapshots, the number of possible snapshot pairs is very large ($\approx 2\times 10^5$), and we wish to omit both early snapshots (because, as we show below, the simulations have not yet settled to steady-state at early times) and snapshots where the lag is so large that our de-rotation procedure becomes questionable. We therefore limit ourselves to considering comparisons between a later snapshot from $t = 500 - 600$ Myr at intervals of 10 Myr, and an earlier snapshot that is $1-50$ Myr prior to that. This gives us a total of 11 samples at each possible lag from $\tau = 1 - 50$ Myr. We take the time correlation coefficient $\xi_a(\tau)$ to be the mean value of these 11 samples, and the uncertainty to be the variance of these samples.

As with the spatial autocorrelation, we fit the measured time-correlation data to the functional form predicted by \citetalias{KT18},
\begin{equation}
\xi(t) =  \frac{\ln\left(1 + 2\phi^2-\phi^2 t/t_\mathrm{corr}\right) - \ln\left(1+\phi^2 t/t_\mathrm{corr}\right)}{\sqrt{\ln \left( 1 + 2 \phi^2 \right) \ln \left( 1 + 2 \phi^2 - 2 \phi^2 t/t_\mathrm{corr} \right) }},
\label{eq:kt18-time}
\end{equation}
where $t_\mathrm{corr}$ is the correlation time to be fit. As with the spatial correlation, we fit this model to the measured time correlations using MCMC sampling. We adopt flat priors in $\log t_\mathrm{corr}$ and $\log \phi$ for $\phi\in[1,100]$ and $t_\mathrm{corr}\in[50,1.5\times10^4]$ Myr, and a Gaussian log-likelihood function $\log\mathcal{L} = -\sum_n [\xi_a(\tau_n) - \xi_\mathrm{mod}(\tau_n)]^2/2\sigma_n^^2$, where the sum is over our $n = 11$ time lag measurements, $\xi_a(\tau_n)$ and $\sigma_n$ are the mean correlation and variance measured at lag $\tau_n$, and $\xi_\mathrm{mod}(\tau_n)$ is the model prediction given by \autoref{eq:kt18-time} for a given pair $(\phi, t_\mathrm{corr})$. We gain using \textsc{emcee} to conduct MCMC sampling \citep{Foreman-Mackey13a}, using 200 walkers and 1000 steps, with first 100 steps discarded for burn-in and the chains thinned by a factor of 5. The MCMC posteriors show a strong covariance between $\phi^2$ and $t_\mathrm{corr}$, motivating the use of the derived timescale $T_\mathrm{corr}\equiv t_\mathrm{corr}/\phi^2$, which is much less covariant with $\phi$; we evaluate this quantity directly from the MCMC samples to preserve covariance. In what follows we will report both $t_\mathrm{corr}$ and $T_\mathrm{corr}$.

\subsection{Element-element cross-correlation}
\label{ssec:analysis_crosscor}

\begin{figure*}
    \centering
    \includegraphics[width=1\textwidth]{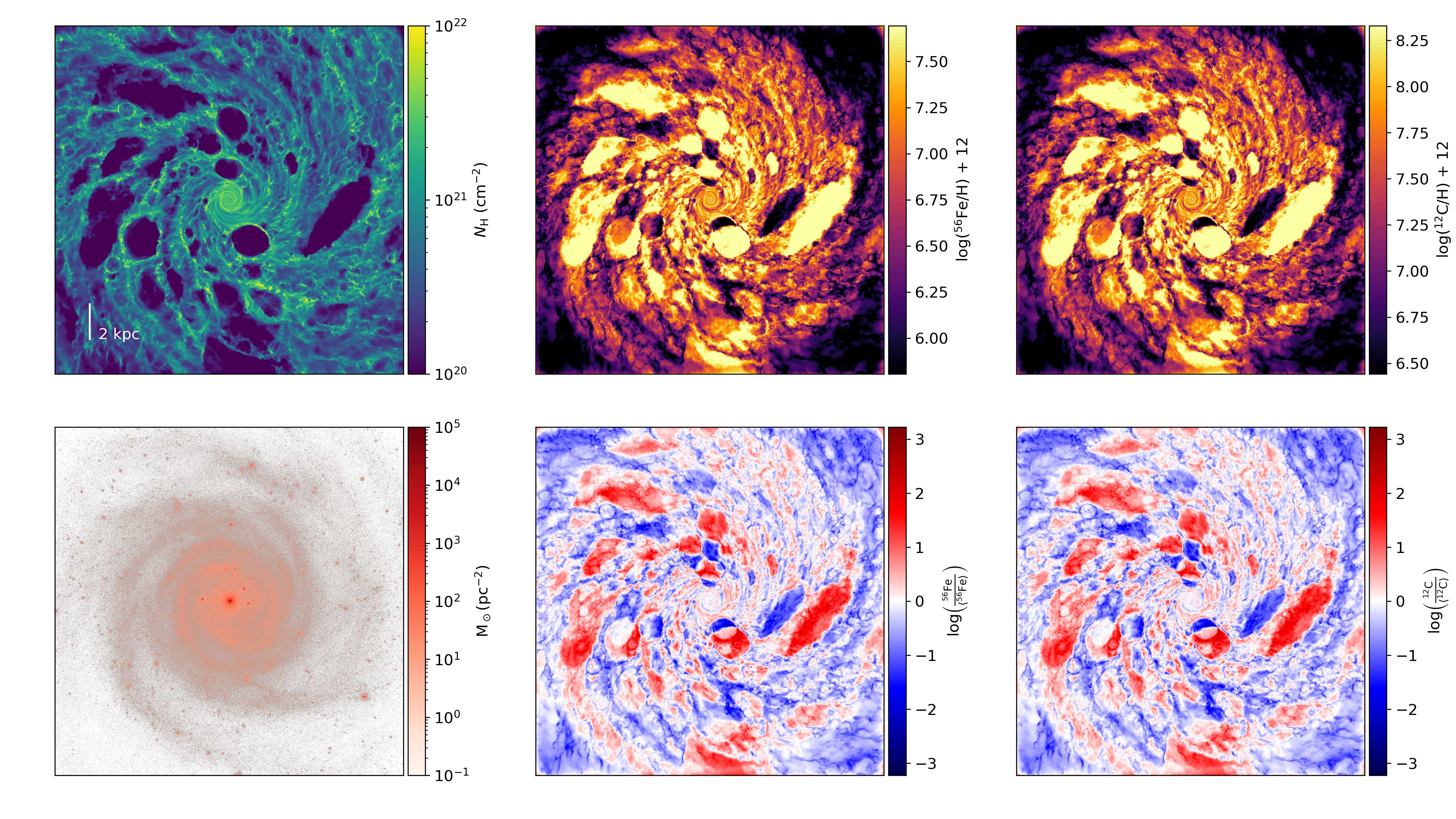}
    \caption{Overview of the RoSE simulation at $t=600$ Myr. Top row: (a) hydrogen column density $N_\mathrm{H}$ (cm$^{-2}$); (b) $^{56}$Fe abundance $\log(^{56}\mathrm{Fe}/\mathrm{H})+12$; (c) $^{12}$C abundance $\log(^{12}\mathrm{C}/\mathrm{H})+12$. Bottom row: (d) stellar surface density $\Sigma_\star$ (M$_\odot$ pc$^{-2}$); (e) $^{56}$Fe fluctuation field $m'({^{56}\mathrm{Fe}})$; (f) $^{12}$C fluctuation field $m'({^{12}\mathrm{C}})$. Each panel covers a 20 kpc $\times$ 20 kpc field centred on the galaxy; a 2 kpc scale bar is shown in panel (a). Note that all abundances shown are measured in the gas rather than in the stars.}
    \label{fig:overview}
\end{figure*}

To measure the degree of similarity between the distributions of different elements, we make use of the cross-correlation. We can define elemental cross-correlations for both the gas and for stars, consistent with the two primary ways of measuring the cross correlation in observations. For gas, we employ the two-point cross-correlation formula from \citetalias{Zhang2025} (their equation 3), at zero lag, which we reproduce here for convenience:
\begin{equation}
    \xi_{a,b} = \left(\frac{1}{\sigma_a \sigma_b}\right)\frac{1}{N_n} 
    \sum_i m'_{a,i} m'_{b,i}.
    \label{eq:corr_implementation}
\end{equation}
Here $m'_{a,i}$ and $m'_{b,i}$ are the metal fluctuation fields for elements $a$ and $b$ in pixel $i$, the sum runs over all $N_n$ pixels, and $\sigma^2_{a,b} = \langle (m'_{a,b})^2\rangle$ are the variances of fields $a$ and $b$ across the whole map.

For stars, we focus on coeval stellar populations, and compute mass-weighted Pearson correlations between elemental abundances in narrow age-selected groups. For each snapshot at time $t$, we bin stars by their formation time (birth epoch) using 1 Myr bins: stars born during $[t_\mathrm{form}, t_\mathrm{form}+1\,\mathrm{Myr})$ constitute one age cohort. This 1 Myr bin width ensures that all stars within a group formed from ISM gas with nearly identical metallicity distributions, enabling a direct measurement of how element-to-element abundance scatter reflects the local ISM composition at their shared birth epoch. For element pair $(a,b)$ and each age bin, we define the correlation as
\begin{equation}
    \rho_{a,b}^{\star} = \frac{\sum_s w_s\,(X_{a,s}-\mu_a)(X_{b,s}-\mu_b)}{\sqrt{\sum_s w_s\,(X_{a,s}-\mu_a)^2}\,\sqrt{\sum_s w_s\,(X_{b,s}-\mu_b)^2}},
    \label{eq:crosscorr_stars}
\end{equation}
where the sum runs over all stars $s$ in a given age bin, $w_s$ is the mass of star $s$, $X_{a,s}$ and $X_{b,s}$ are the abundances of elements $a$ and $b$ in star $s$, $\mu_a = \big(\sum_s w_s X_{a,s}\big)/\big(\sum_s w_s\big)$ is the weighted mean abundance of element $a$ over all stars in the bin, and similarly for element $b$. We compute $\rho_{a,b}^{\star}$ for each 1 Myr age bin across the full simulation epoch (1--600 Myr), yielding a time series of stellar cross-correlations that reflects how the ISM's nucleosynthetic imprint evolves and is preserved in successive stellar generations.

\section{Results}

We first present a qualitative overview of the simulation outcome in \autoref{ssec:results_qualitative}. We then present results for the spatial and temporal correlations of elements in \autoref{ssec:results_spatial} and \autoref{ssec:results_temporal}, and finally examine elemental cross-correlations in \autoref{ssec:results_crosscor}.

\subsection{Qualitative overview}
\label{ssec:results_qualitative}

\autoref{fig:overview} shows the state of the simulation at $t=600$ Myr. The top row displays hydrogen column density $N_\mathrm{H}$ (left), total $^{56}$Fe abundance (i.e., $^{56}$Fe contributed by all nucleosynthetic channels; middle), and total $^{12}$C abundance (right). The bottom row shows stellar surface density (left) and gas metallicity fluctuation maps $m'$ for the same isotopes of Fe (middle) and C (right). The gas distribution exhibits well-developed spiral structure with peak column densities exceeding $10^{22}\,\mathrm{cm}^{-2}$ in star-forming regions. On large scales, metallicity exhibits a strong correlation with gas distribution, such that regions of low gas density often display higher metallicity. This arises because the cavities blown out by supernova explosions are invariably more metal-rich. Metal abundance maps also reveal small r-scale inhomogeneities superimposed on these large-scale structures, reflecting the discrete injection events and subsequent turbulent mixing that characterize the ISM enrichment process. The stellar disc traces the gas spiral arms with a time delay corresponding to stellar drift. Metallicity fluctuations (panels e and f) quantify departures from radial profiles, highlighting coherent metal-rich and metal-poor patches on scales of a few hundred parsecs -- the signatures of individual star formation episodes and feedback events that are the subject of our correlation analysis.

\begin{figure*}
    \centering
    \includegraphics[width=1\textwidth]{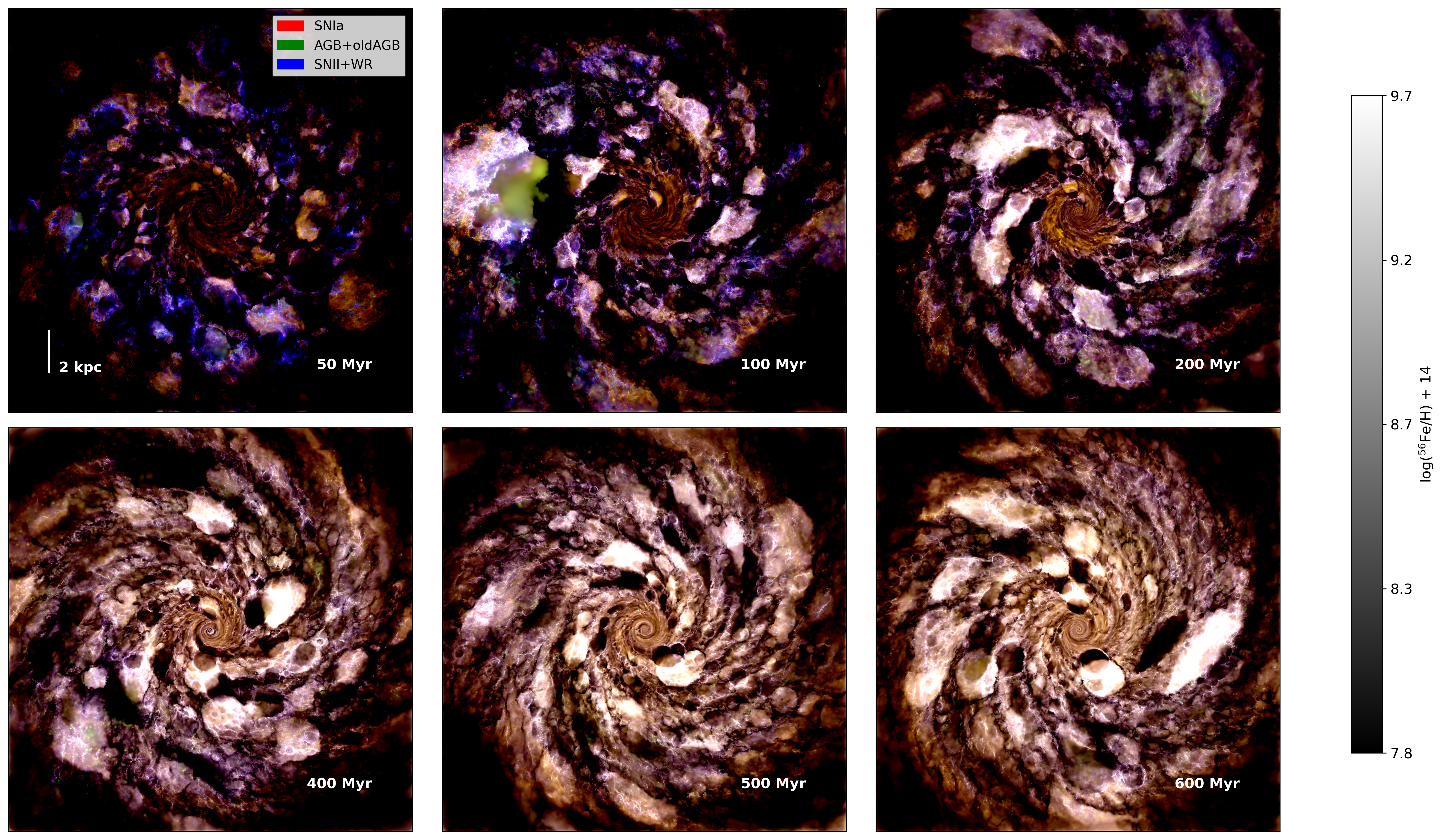}
    \caption{Maps of $^{56}$Fe abundance in interstellar gas decomposed into three colour channels corresponding to different nucleosynthetic origin sites: SNIa (red), AGB stars (green, including both short- and long-timescale AGB contributions), and SNII+WR winds (blue). Each channel's intensity encodes its fractional abundance relative to hydrogen, as indicated in the colourbar. Each panel is $20\;\mathrm{kpc}\times 20\;\mathrm{kpc}$ in size, and shows the simulation state at times from 50 to 600 Myr as indicated by the labels; the colour scale is the same in all six panels, and is set so the range from zero to maximum intensity corresponds to the 10--90th percentiles of the total $^{56}$Fe/H field at $t=600$ Myr.}
    \label{fig:fe56_rgb}
\end{figure*}

\autoref{fig:fe56_rgb} illustrates how spatial distributions for the contributions from different nucleosynthetic sources vary, using $^{56}$Fe as an example. In this figure, we use the red, green, and blue channels to represent $^{56}$Fe produced by SNIa, AGB stars (both young and old), and SNII and WR stars, respectively. 
At early times ($t=50$ Myr), discrete supernova-driven bubbles are clearly visible as blue (SNII+WR) and red (SNIa) structures, their sizes and morphologies reflecting the local gas density -- denser nuclear regions host more compact and numerous bubbles, while lower-density outer regions show larger, more extended cavities. Individual events imprint distinct signatures: the early Type II and Type Ia explosions carve out structures that persist as recognizable features throughout hundreds Myr evolution.

With advancing time, turbulent mixing progressively blends the three colour channels, and by $t \approx 400$ Myr the colour distribution seems to have settled into a rough statistical steady-state in which the large-scale metal distribution has become relatively homogeneous. However, even at the final epoch ($t=600$ Myr), the RGB decomposition retains significant structure at small scales, revealing that chemical differentiation persists in patches corresponding to recent star-forming regions and their feedback-driven bubbles. This persistence of such structures is the natural consequence of continuous nucleosynthetic injection: as older metal-enriched regions mix and homogenize, newly-injected metals from ongoing star formation create fresh spatial gradients and correlations. 

A critical observation we can already make from \autoref{fig:overview} and \autoref{fig:fe56_rgb} is the radial dependence of the efficiency with which gas is mixed in our simulation. Within the inner region ($r \lesssim 5$ kpc), the source-decomposed colour distinctions are less pronounced than in the outer disc, indicating more efficient local mixing in the simulated gas. In contrast, the outer disc retains more pronounced source-to-source variation, reflecting both lower turbulent activity and the geometrically larger separation of injection sites relative to the diffusion length scale. This radial gradient in chemical homogeneity has important consequences for interpreting abundance patterns in different galactic environments.

\begin{figure}
    \centering
    \includegraphics[width=1\columnwidth]{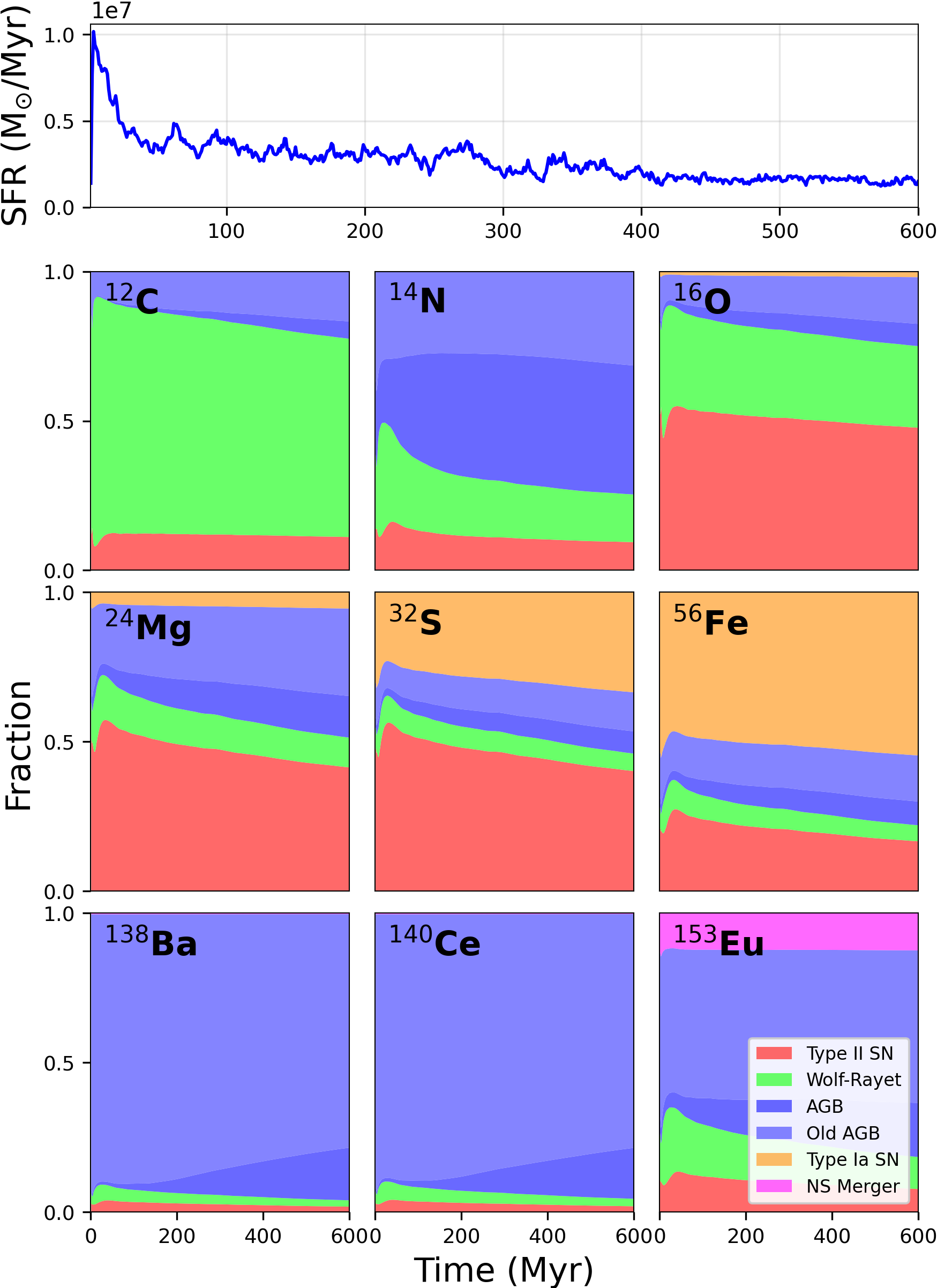}
    \caption{The nine lower panels show the fractional cumulative contribution of each nucleosynthetic channel to each of nine tracked isotopes as a function of time; for comparison the top panel shows the star formation rate (SFR) over the same time period. The stacked horizontal bars in each of the nine element panels show the fraction of the mass of the indicated element that was produced up to that time by each of the six channels indicated in the legend. Note that as the SFR approaches equilibrium at later times, the elemental fractions also stabilize, reflecting the steady-state balance between different nucleosynthetic channels.}
    \label{fig:source_fractions}
\end{figure}

Finally, \autoref{fig:source_fractions} shows the fractional contribution of each modelled nucleosynthetic channel to each of the nine isotopes we track as a function of time in the simulation. As expected, the composition of each element reflects the nucleosynthetic pathways encoded in the stellar yields: WR winds produce most C, SNII dominates the $\alpha$ elements (O, Mg, S), SNIa contributes the majority of iron-peak elements (Fe), AGB stars contribute significantly to neutron-capture elements (Ba, Ce) and moderately to carbon and nitrogen, and the rare neutron star merger channel leaves an imprint primarily on the $r$-process element Eu, but is negligible for all other elements. However, we can also see an important artefact over the first $\approx 50$ Myr of the simulation: a transient enhancement of WR and SNII source fractions. This feature reflects our progressively-increasing resolution initialisation procedure as described in \autoref{ssec:initcon}: when we increase the resolution, this leads to a temporary burst of star formation, which in turn leads to a transient injection of SNII and WR elements a short time later. We will therefore generally ignore the first $\approx 100$ Myr of the simulation in the analyses the follow.

\subsection{Spatial scales and elemental clustering}
\label{ssec:results_spatial}

\begin{figure}
    \centering
    \includegraphics[width=1\columnwidth]{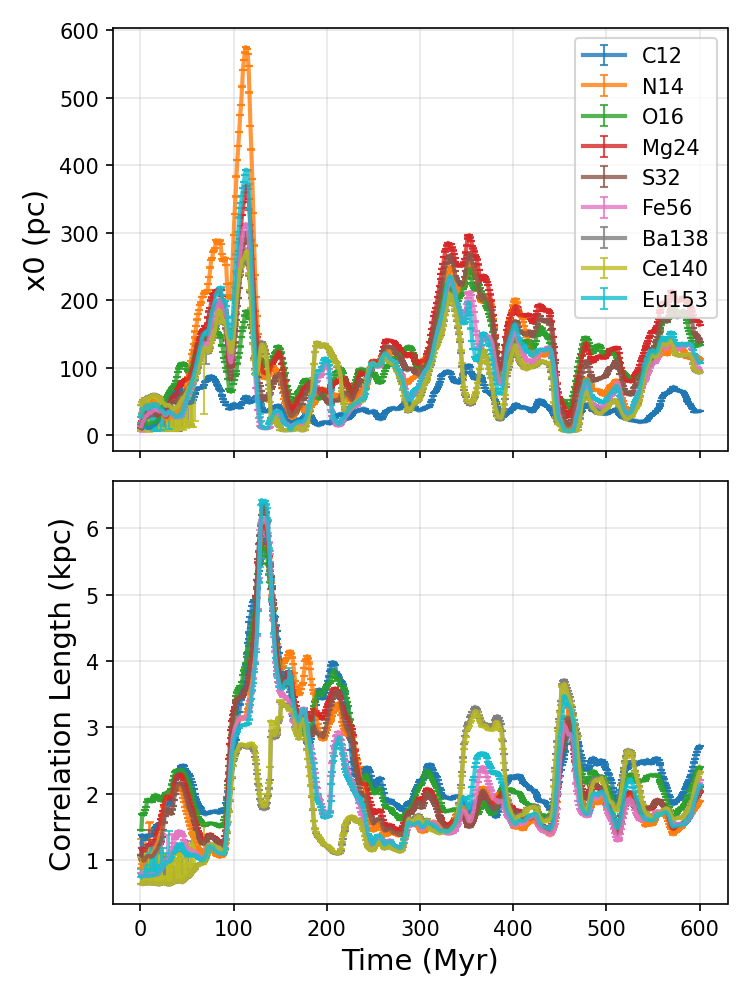}
    \caption{Autocorrelation parameters versus time evolution for all isotopes, computed using the total metal field fluctuation field summed over all nucleosynthetic channels. The top panel shows the injection width $x_0$, the bottom shows the correlation length $l_\mathrm{corr}$. Each coloured curve corresponds to one isotope, as indicated in the legend. Points show the 50th percentile value and error bars indicate the 5–95 percentile ranges from the MCMC fits (see \autoref{ssec:analysis_acorr}).}
    \label{fig:auto_all_multi}
\end{figure}

We next quantify the spatial structure and its time evolution via the autocorrelation fits described in \autoref{ssec:analysis_acorr}. In \autoref{fig:auto_all_multi}, we plot our parametric fits to the injection scale $x_0$ and correlation length $l_\mathrm{corr}$ for all elements versus time, summing over all nucleosynthetic sources. We see that, after a transient period in the first $\approx 200$ Myr of the simulation, the injection scale fluctuates around tens of parsecs and the correlation length oscillates around 1--3 kpc for all elements. The transient spikes visible in these plots correspond to periods of upward fluctuations in the star formation rate (compare \autoref{fig:auto_all_multi} to the upper panel in \autoref{fig:source_fractions}), during which newly-injected supernova-enriched material temporarily compresses the spatial scale of the metal distribution. 

The correlation lengths $l_\mathrm{corr}$ and injection scales $x_0$ that we find for $^{16}$O, the element whose spatial structure is most easily measurable in the gas phase, are in good agreement with the results of extragalactic studies. For example, studies of large samples of galaxies ($\gtrsim 100$) with star formation rates $\dot{M}_*\sim 1$ M$_\odot$ yr$^{-1}$, comparable to our simulated galaxy, are observed to have $l_\mathrm{corr} \sim 1$ kpc \citep[e.g.,][]{2019ApJ...887...80K, 2020MNRAS.499..193K, 2021MNRAS.508..489M, Li2021, Li_2022, 2022MNRAS.509.1303W}, and higher-resolution follow-ups that have measured $x_0$ typically find values $\sim 50$ pc \citep[e.g.,][]{Li_2022}, consistent with the results shown in \autoref{fig:auto_all_multi}. More subtly, by comparing metallicity correlations in galaxies with and without active nuclei, \citet{2024arXiv240704252L} conclude that correlation lengths are not primarily determined by galaxy assembly histories or other processes acting over cosmological timescales, but instead reach an equilibrium determined primarily by galaxies' instantaneous star formation rates on much shorter timescales; this is consistent with our finding that that galaxy correlation lengths settle to rough steady state in only a few hundred Myr.

This agreement should be interpreted with some caution because nebular emission lines measure oxygen remaining in the gas phase, whereas RoSE does not partition oxygen between gas and dust. Observations of H\,{\sc ii} regions indicate that a non-negligible fraction of oxygen is depleted onto grains and that the depletion increases with metallicity \citep{Peimbert2010Dust}. A spatially uniform depletion would mainly shift the abundance zero-point and leave our normalised correlation statistics unchanged, but variations with metallicity or local ISM conditions could introduce or suppress spatial fluctuations in gas-phase oxygen and thereby bias the inferred $l_\mathrm{corr}$ and $x_0$ relative to the total-oxygen values predicted by RoSE. We return to this limitation in \autoref{ssec:caveats}.

If we now examine the relationship between elements rather than focusing on $^{16}$O alone, a striking feature of \autoref{fig:auto_all_multi} is the clear \emph{elemental clustering} in parameter space. Carbon ($^{12}$C), primarily sourced from WR winds, consistently exhibits the lowest injection scale and highest correlation length throughout the simulation. In contrast, the core-collapse products ($^{16}$O, $^{24}$Mg, $^{32}$S from SNII) show intermediate values that vary nearly identically in time. Iron ($^{56}$Fe), dominated by the delayed SNIa channel, traces a distinct path offset to larger injection scale and shorter correlation length, reflecting its longer delay-time distribution. The neutron-capture elements (Ba, Ce) and the $r$-process element Eu form yet another cluster, characterized by slower initial growth in $x_0$ and sustained high correlation lengths. This element-by-element partitioning of the parameters directly reflects the differences in delay-time distribution and injection environment between channels. 

\begin{figure}
    \centering
    \includegraphics[width=1\columnwidth]{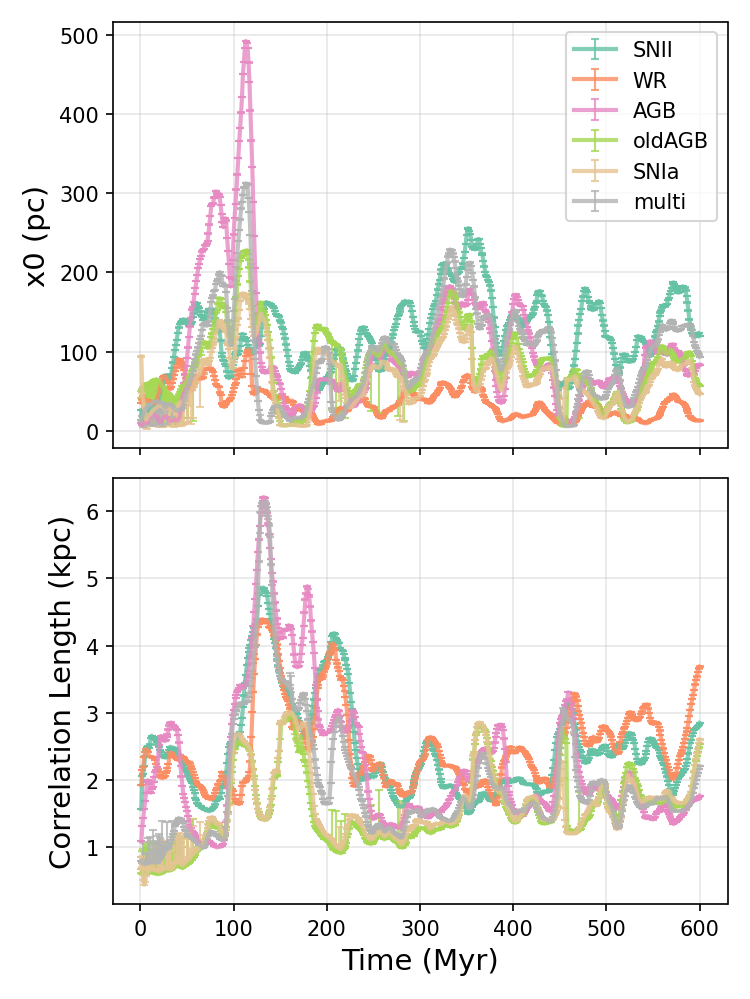}
    \caption{Same as \autoref{fig:auto_all_multi}, but now showing parameters for fits to just the $^{56}$Fe fluctuation field decomposed by nucleosynthetic source -- SNII, WR, AGB, oldAGB, SNIa -- as indicated in the legend. We also show the results for the total $^{56}$Fe field (labelled ``multi'') for comparison; this curve is the same as the $^{56}$Fe shown in \autoref{fig:auto_all_multi}.}
    \label{fig:auto_fe_sources}
\end{figure}

To confirm that the differences between elements are indeed driven mainly by their different origin sites, in
\autoref{fig:auto_fe_sources} we disentangle the contributions of individual nucleosynthetic channels to the spatial structure of $^{56}$Fe, presenting parametric fits to the part of the $^{56}$Fe distribution contributed by each of the five nucleosynthetic channels that make a non-zero contribution to it (all except NSM), alongside the total (multi) field. It is clear from this figure that, once the galaxy settles into near steady-state after a few hundred Myr, the parameters for the \emph{total} field of a given element (grey lines) closely track those of its dominant nucleosynthetic source (SNIa; beige line). While the figure shows results of $^{56}$Fe only, we find the same qualitative trend for other isotopes. Thus we conclude that the spatial statistics of the distribution of an element in galaxies is predominantly determined by its main product nucleosynthetic origin channel.

\subsection{Temporal correlations}
\label{ssec:results_temporal}

\begin{figure}
    \centering
    \includegraphics[width=1\columnwidth]{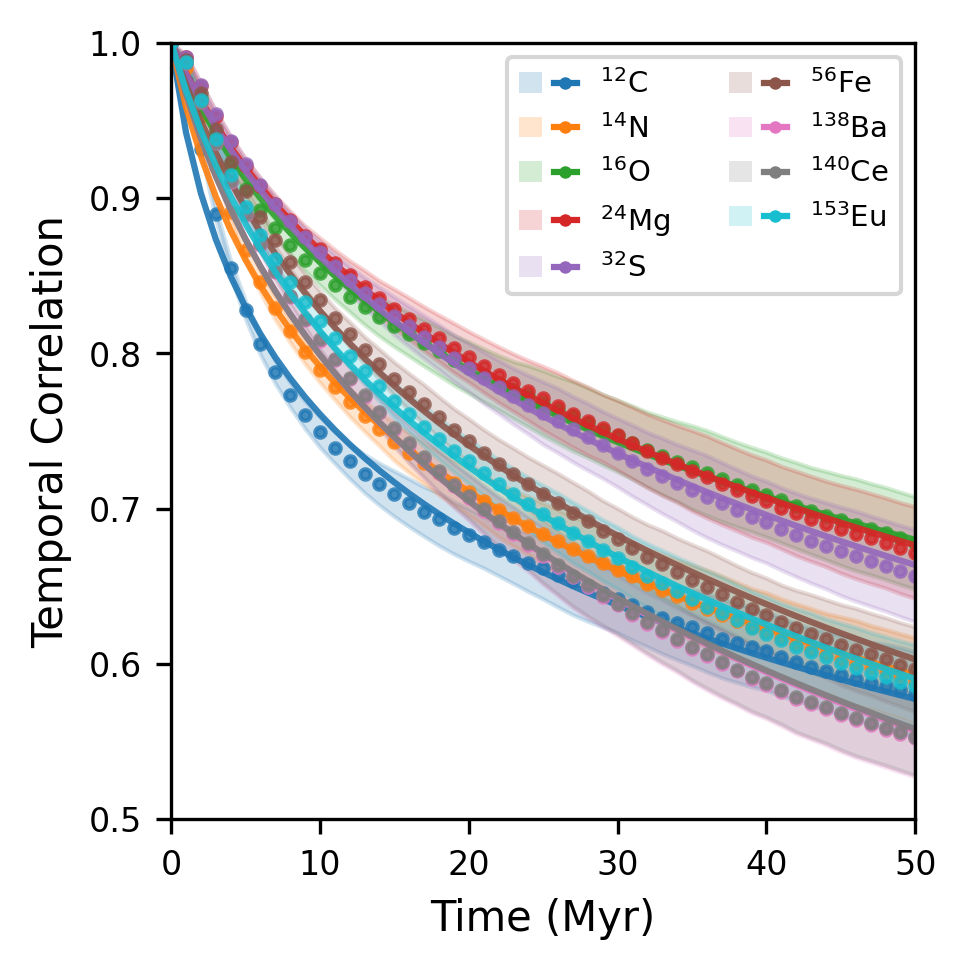}
    \caption{Data points and shaded bands show the measured mean temporal correlation and its variance as a function of time lag, with values computed via the procedure outlined in \autoref{ssec:analysis_tcorr}. Solid lines show best-fitting \citetalias{KT18} models, with fit parameters shown in \autoref{tab:time_corr_params}.}
    \label{fig:time_correlation_all_isotopes}
\end{figure}

We next examine the temporal correlations of the metal field, which we compute following the procedure outlined in \autoref{ssec:analysis_tcorr}. We show the measured correlation as a function of time lag for each element in \autoref{fig:time_correlation_all_isotopes}, together with the corresponding best-fitting \citetalias{KT18} model (\autoref{eq:kt18-time}).\footnote{Careful readers may notice some difference from the results in \citetalias{Zhang2025}, even for elements that were included in that paper. This was due to a coding error in \citetalias{Zhang2025}, which was discovered in the course of preparing this work, whereby we computed the time correlation on the non-logged rather than the logged metallicity fluctuation field. That error has been corrected in the present paper.} It is clear that the functional forms predicted by the models provide a good description of the data. We report our fit parameters and their uncertainties in \autoref{tab:time_corr_params}.

Examining the table and figure, we again see a trend between temporal correlation and nucleosynthetic origin site. The elements whose production is dominated by SNII -- O, Mg, and S -- cluster tightly together and have the longest correlation time. The elements produced by SNIa, AGB stars, and NS mergers -- N, Fe, Ba, Ce, and Eu -- form a clearly distinct group with a shorter correlation time than the SNII elements. Finally, the one element whose production is largely via WR stars -- C -- stands out as quite distinct from all other elements in that its correlation drops rapidly in time, but then flattens considerably. 

\begin{figure}
    \centering
    \includegraphics[width=\columnwidth]{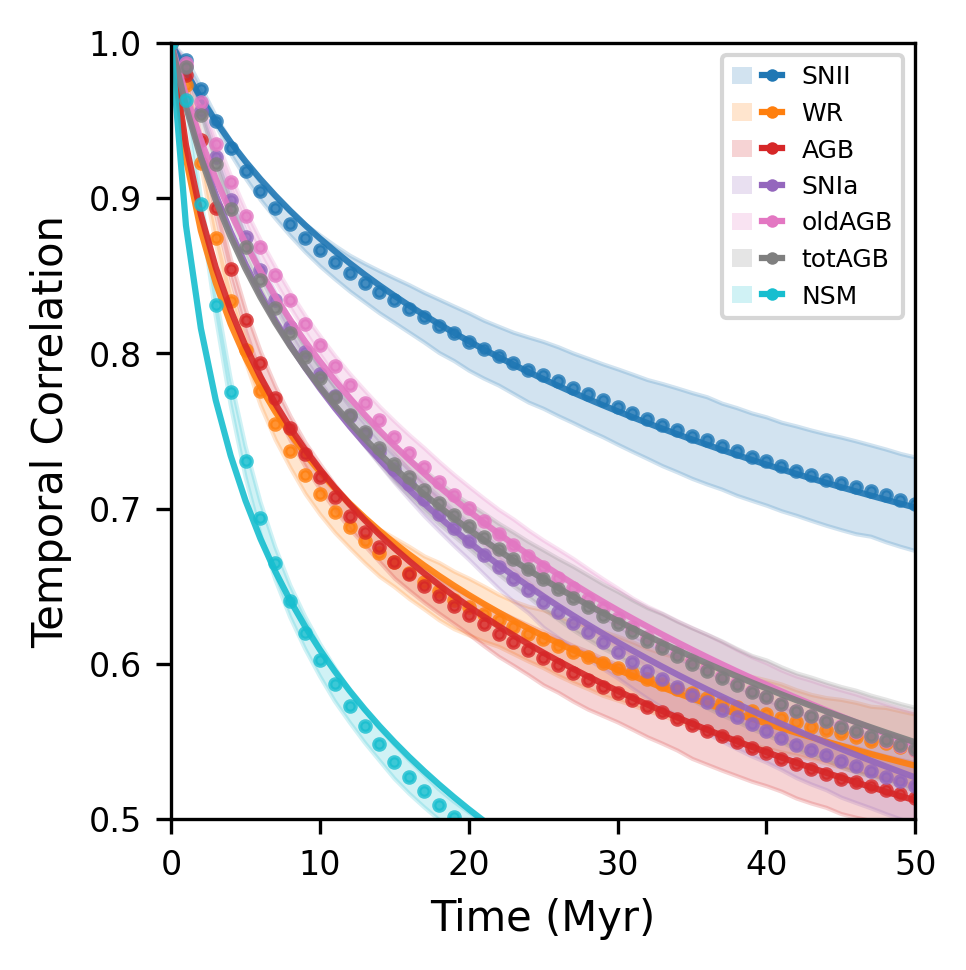}
    \caption{Same as \autoref{fig:time_correlation_all_isotopes}, but now showing results for Fe only, decomposed by the contribution to Fe from different nucleosynthetic origin sites. Parameters for the best-fitting \citetalias{KT18} models (solid lines) are given in \autoref{tab:time_corr_sources}.}
    \label{fig:tab:time_corr_components}
\end{figure}

\autoref{fig:tab:time_corr_components} confirms that these differences can indeed be attributed to nucleosynthetic origin site. In analogy to \autoref{fig:auto_fe_sources}, in this figure we show the temporal correlation calculated separately for each of the nucleosynthetic sources that produce Fe; the results for other elements are again similar. We report \citetalias{KT18} model fit parameters in \autoref{tab:time_corr_sources}. In this figure and table the differences in correlation time for elements produced at different astrophysical sites becomes extremely clear. Production by SNII, which tends to produce large, long-lived bubbles of enrichment, yields the longest time correlation, while NSMs, which are rare events that do not produce large bubbles because they inject limited energy, decorrelate extremely rapidly. Other sources are intermediate.

\begin{table}
    \centering
    \caption{\citetalias{KT18} model fit parameters from time-correlation analysis for all tracked isotopes using the total (multi-source) metal field. $\phi$ is the dimensionless shape parameter; $t_\mathrm{corr}$ (Myr) is the correlation time; $T_\mathrm{corr} \equiv t_\mathrm{corr}/\phi^2$ (Myr) is the characteristic temporal decorrelation timescale. Uncertainties are 16--84 percentile ranges from the MCMC posterior.}
    \label{tab:time_corr_params}
    \begin{tabular}{lccc}
        \hline
        Isotope & $\phi$ & $t_\mathrm{corr}$ (Gyr)\\
        \hline
        $^{12}$C   & $41^{+33}_{-22}$ & $2.8^{+3.2}_{-1.7}$\\
        $^{14}$N   & $25^{+32}_{-14}$ & $1.8^{+3.6}_{-1.2}$\\
        $^{16}$O   & $24^{+24}_{-14}$ & $3.5^{+5.8}_{-2.4}$\\
        $^{24}$Mg  & $15^{+22}_{-9}$ & $1.8^{+4.7}_{-1.3}$\\
        $^{32}$S   & $13^{+22}_{-8}$ & $1.5^{+4.1}_{-1.0}$\\
        $^{56}$Fe  & $12^{+21}_{-7}$ & $8.4^{+2.4}_{-0.5}$\\
        $^{138}$Ba & $11^{+17}_{-6}$ & $0.6^{+1.3}_{-0.3}$\\
        $^{140}$Ce & $11^{+18}_{-6}$ & $0.6^{+1.4}_{-0.4}$\\
        $^{153}$Eu & $13^{+23}_{-8}$ & $0.9^{+2.3}_{-0.6}$\\
        \hline
    \end{tabular}
\end{table}

The quantitative agreement between spatial and temporal KT parameters reinforces the self-consistency of the diffusion-injection framework: elements with larger spatial correlation lengths (e.g., Ba, Ce, Eu) also exhibit longer temporal decorrelation times, while elements with shorter spatial scales (O, Mg, S) decorrelate more rapidly in time. This consistency validates the physical picture wherein nucleosynthetic delay times set both the characteristic injection scale (through the spatial extent of parent stellar populations) and the mixing timescale (through the interval between successive injections at a given location).

\begin{table}
    \centering
    \caption{Same as \autoref{tab:time_corr_params}, but now showing the results of \citetalias{KT18} model fits to the temporal correlation of the $^{56}$Fe metal field decomposed into contributions from individual nucleosynthetic sites.}
    \label{tab:time_corr_sources}
    \begin{tabular}{lccc}
        \hline
        Component & $\phi$ & $t_\mathrm{corr}$ (Gyr)\\
        \hline
        SNII    & $23^{+21}_{-13}$ & $3.8^{+5.9}_{-2.7}$\\
        WR      & $41^{+33}_{-21}$ & $2.0^{+2.1}_{-1.1}$\\
        AGB     & $24^{+29}_{-13}$ & $1.0^{+1.5}_{-0.6}$\\
        SNIa    & $11^{+17}_{-6}$ & $0.5^{+1.0}_{-0.3}$\\
        oldAGB  & $10^{+16}_{-5}$ & $0.5^{+1.1}_{-0.3}$\\
        totAGB  & $15^{+22}_{-8}$ & $0.8^{+1.6}_{-0.4}$\\
        NSM    & $18^{+21}_{-8}$ & $0.2^{+0.3}_{-0.1}$\\
        \hline
    \end{tabular}
\end{table}

\subsection{Elemental cross-correlations}
\label{ssec:results_crosscor}

\begin{figure}
    \centering
    \includegraphics[width=1\columnwidth]{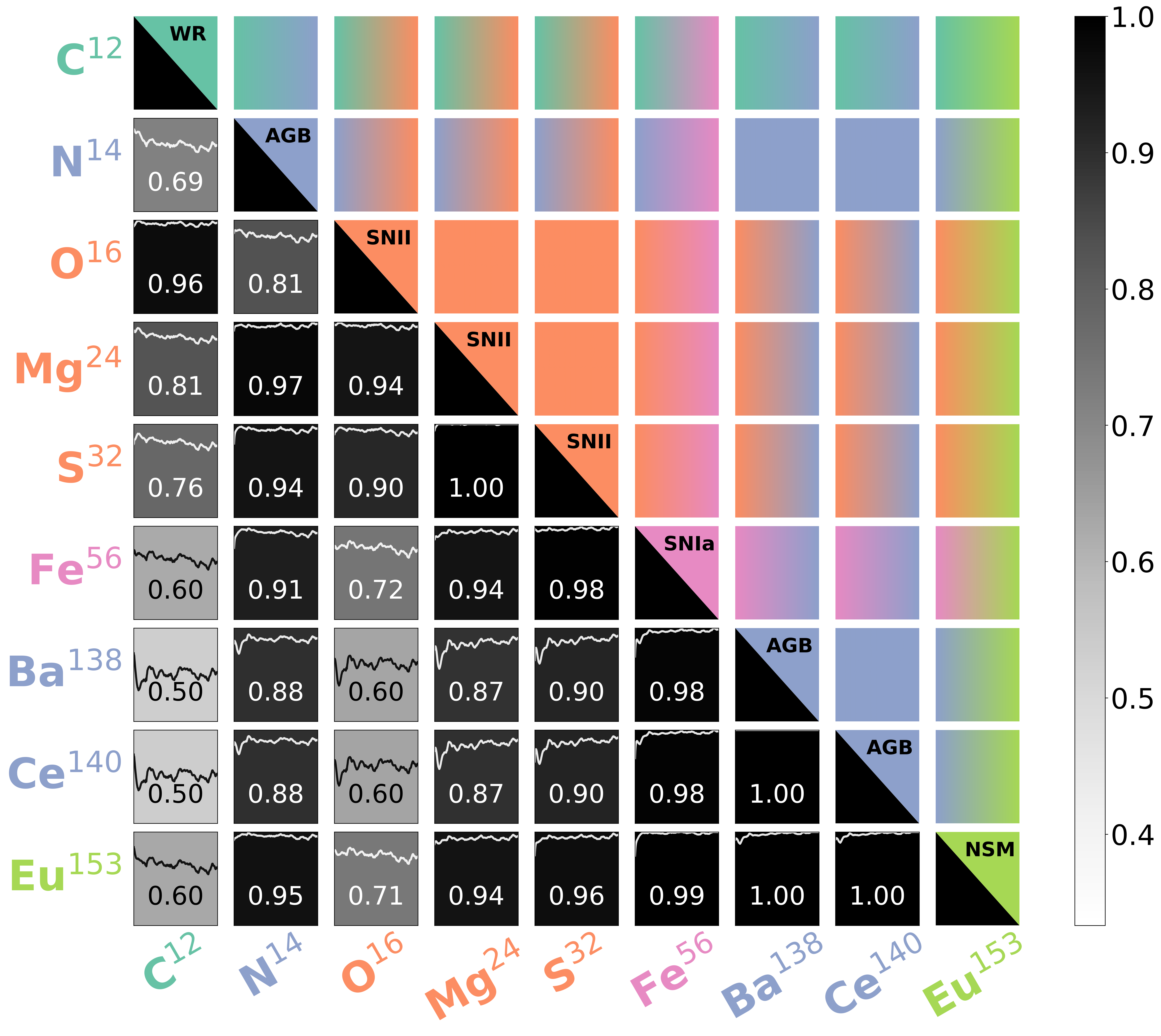}
    \caption{Gas-phase cross-correlation matrix for all nine tracked isotopes spanning the full 1--600 Myr simulation. Each cell $(i,j)$ shows the time evolution of the two-point cross-correlation coefficient for element pair $i$ versus $j$ (plotted as a white curve), with the background color (red colormap) and numerical value indicating the time-averaged correlation strength over the final 100 Myr quasi-equilibrium epoch. The lower-triangle layout shows all unique pairs; the diagonal is unity (self-correlation). To guide the eye, the shaded upper-triangle highlights the dominant production channel of each element (except Eu as it is the only element has significant NSM contribution) with the colours indicating the different channels (see legend along the diagonal).  Note that elements sharing the same dominant source (e.g., O, Mg, S from SNII) show consistently high correlation; pairs mixing sources exhibit reduced correlation reflecting their distinct delay-time distributions.}
    \label{fig:cross_corr_gas_2d}
\end{figure}

Our final analysis examines the cross-correlations between different elements in both the gas and stellar phases, computed as described in \autoref{ssec:analysis_crosscor}. \autoref{fig:cross_corr_gas_2d} shows the former; in this diagram, the numerical value and colour for each element-element pair show the mean cross-correlation averaged over the last 100 Myr of the simulation, while white lines within each panel show the cross-correlation as a function of time, with the horizontal axis running over the full 600 Myr duration of the simulation and the vertical axis going from 0 to 1. We see that elements sharing the same dominant source (e.g., O, Mg, S -- all from SNII) show very strong correlations, while pairs mixing sources (e.g., WR-dominated C with AGB-dominated Ba, Ce) exhibit smaller correlations. This is a direct signature of the difference in injection sites: the two distinct nucleosynthetic sources create spatially offset patterns, which are then progressively erased over timescales of the correlation time. The final level of correlation is set by the competition between ongoing injection and erasure by mixing. Consistent with this picture, and with the analytic model of \citet{KT25}, we find that for element pairs from mixed nucleosynthetic sources, the degree of correlation appears to be related to how far apart the two nucleosynthetic channels are in their delay times. The example of C, which is returned on the shortest timescales due to the dominance of WR stars in its production, is illustrative: C correlates best with the SNII elements (O, Mg, S), which have the next-shortest delay time, followed by N from short-lived AGB stars, then Fe from SNIa, and finally the elements produced primarily by old AGB stars and NSM elements (Ba, Ce, Eu), which have the longest delay times. Thus the degree of correlation between two elements appears to be closely linked to how different their delay times are between star formation and element return. 

\begin{figure}
    \centering
    \includegraphics[width=1\columnwidth]{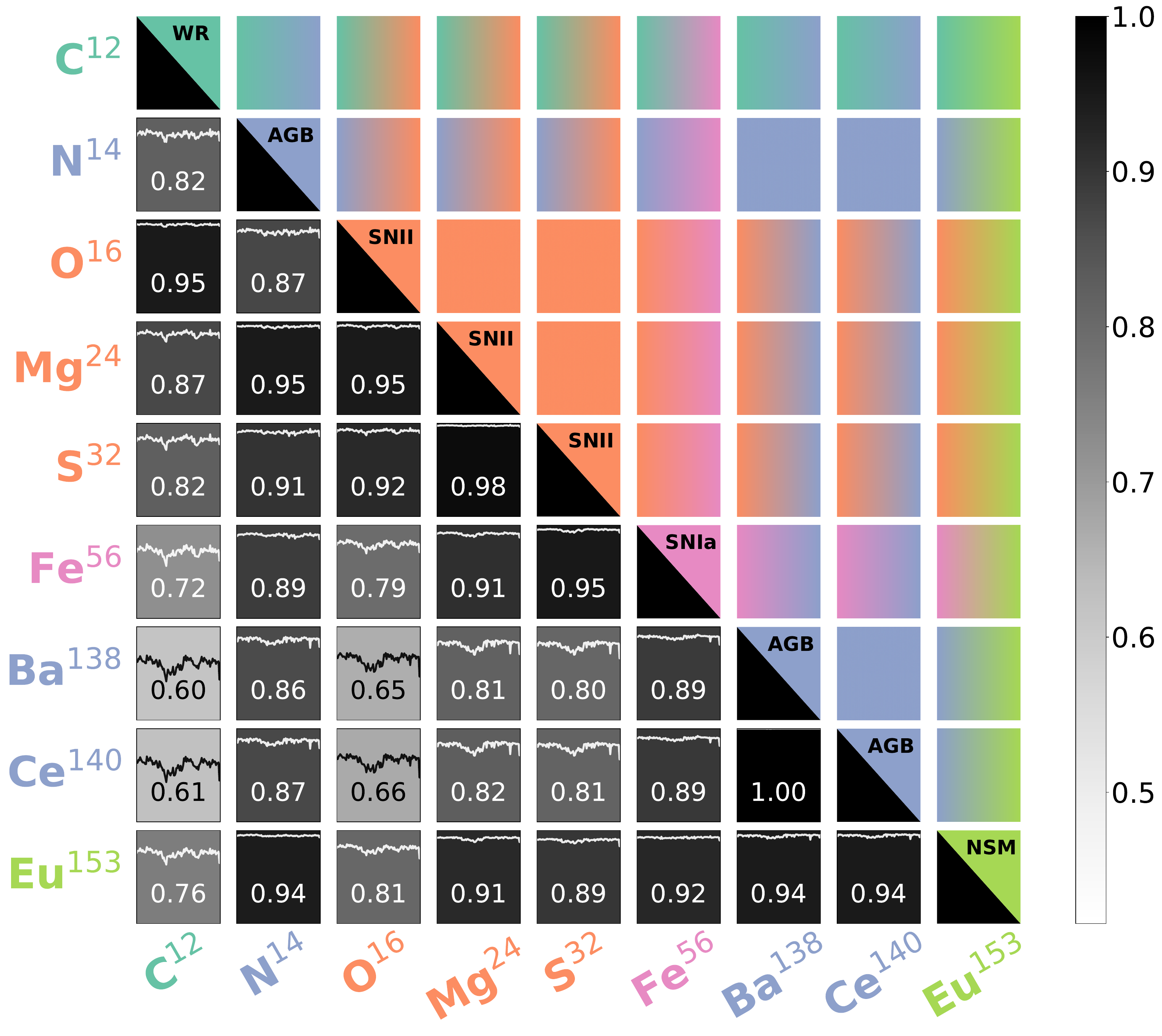}
    \caption{Same as \autoref{fig:cross_corr_gas_2d}, but now showing the cross-correlations in stars. The white lines showing time evolution measure the cross-correlation in stars born at a given time. The correlations are weighted by stellar particle mass following \autoref{eq:crosscorr_stars}, and the curves and reported means cover birth epochs $t=501$--$600$ Myr.}
    \label{fig:stellar_cross}
\end{figure}

\autoref{fig:stellar_cross} presents the corresponding stellar-phase cross-correlations. Each time point as indicated by the white lines in this plot represents the internal correlation for stars born in the same 1 Myr cohort: for example, the correlation at $t=550$ Myr is computed from stars born from 550--551 Myr into the simulation. We see that the overall correlation structure mirrors that observed in the gas phase, demonstrating that the nucleosynthetic source imprints present in the ISM at any given moment are faithfully recorded in the stellar populations formed during that epoch and persist as permanent chemical signatures. However, quantitative differences emerge when comparing gas and stellar correlation strength element-pair-by-element-pair: prompt-source pairs (e.g., C--O from WR+SNII) show stronger correlation in stars than in gas, because stars formed early in the simulation inherited the ISM's high correlation established during the initial rapid enrichment phase, and this signature is preserved indefinitely; conversely, pairs involving delayed sources (e.g., Fe correlations with alpha elements) exhibit higher gas-phase values, reflecting the ongoing SNIa enrichment that continues to modify the ISM but has not yet fully imprinted the cumulative stellar disc. This phase-dependent variation encodes information about the nucleosynthetic progenitor timescales and the time-integrated enrichment history. The stabilisation timescale for stellar cross-correlations is also considerably shorter than decorrelation timescales measured in \autoref{ssec:results_temporal}. This rapid convergence arises because each stellar cohort samples only a narrow time slice of the ISM, avoiding the temporal averaging that broadens the gas-phase correlation timescales; however, the finite delay between metal ejection, gas cooling, and star formation introduces a lag, so stellar cross-correlations reach equilibrium slightly later than their gas counterparts.

\begin{figure}
    \centering
    \includegraphics[width=1\columnwidth]{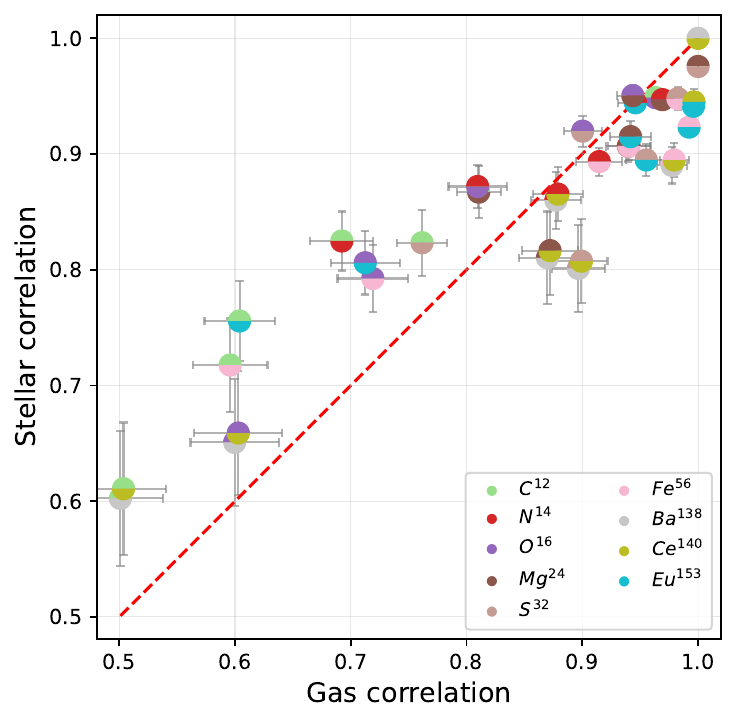}
    \caption{Direct comparison of gas and stellar cross-correlations across element pairs. Each point represents an element pair (colored half-circles indicate the two constituent elements), computed from gas metallicity fluctuations over the time interval $t=501-600$ Myr ($x$-axis) and stellar age bins $0-100$ Myr at the final snapshot ($y$-axis). Error bars denote standard deviations across individual time steps (gas) and age bins (stellar). The red dashed line marks $y=x$ as a reference.}
    \label{fig:gas_vs_star}
\end{figure}

As shown in \autoref{fig:gas_vs_star}, these quantitative comparisons between gas and stellar phase correlations reveal a fundamental principle: element pairs with prompt nucleosynthetic sources show stronger correlations in stars than in gas, because stellar populations preserve the high correlations established during rapid early enrichment; conversely, pairs involving delayed sources show higher gas correlations, reflecting the ongoing enrichment that continues to modify the ISM but has not yet fully imprinted the stellar populations. This phase-dependent behaviour directly encodes information about nucleosynthetic timescales and enrichment histories, motivating a deeper mechanistic analysis in the Discussion section.

\section{Discussion}

Our results demonstrate that numerous aspects of elemental distributions within galaxies -- spatial structure, correlation in time, and element-element cross-correlation in both gas and stars -- appear to be directly traceable to those elements diverse nucleosynthetic origin sites. In this discussion we seek to understand this behaviour in more detail.

\subsection{A simple linear model for cross-correlation}

As a first step, we investigate to what extent a very simple linear model can predict how correlated two elements will be solely based on the amounts by which different nucleosynthetic channels contribute to their production. To this end for each element pair $(e_1, e_2)$ and snapshot, we define
\begin{equation}
    \Delta_k(e_1, e_2) = \left| f_k(e_1) - f_k(e_2) \right|,
\end{equation}
where $f_k(e)$ is the fractional contribution of nucleosynthetic channel $k \in \{\mathrm{SNII, WR, AGB, SNIa, NSM, oldAGB}\}$ to element $e$'s total abundance at that time, i.e., the quantity shown in \autoref{fig:source_fractions}. Thus at each time and for each element pair we obtain a 6-dimensional vector $\boldsymbol{\Delta} = [\Delta_\mathrm{SNII}, \Delta_\mathrm{WR}, \Delta_\mathrm{AGB}, \Delta_\mathrm{SNIa}, \Delta_\mathrm{NSM}, \Delta_\mathrm{oldAGB}]$ that describes how different two elements are in their nucleosynthetic origins; element pairs with very similar origins will have all elements of $\boldsymbol{\Delta}$ close to zero, while for two elements with completely disjoint nucleosynthetic origins the entries in $\boldsymbol{\Delta}$ will sum to two. We calculate $\boldsymbol{\Delta}$ for times $t=501$--600 Myr at 1 Myr intervals.

To test how well $\boldsymbol{\Delta}$ can predict element-element cross correlations, we now consider a simple model of the form
\begin{equation}
    \xi_{e_1,e_2}^\mathrm{pred} = \mathbf{c}^{\mathsf{T}} \boldsymbol{\Delta} + b,
    \label{eq:linear_regression}
\end{equation}
where $\xi_{e_1,e_2}^\mathrm{pred}$ is the predicted cross-correlation between elements $e_1$ and $e_2$ (in either the stellar or gas phase), and $\mathbf{c} \in \mathbb{R}^6$ and $b$ are the source-wise regression coefficients and intercept, which we leave as free parameters to be fit. Before proceeding to fit the model, however, we improve the underlying data by introducing three ``pure'' basis elements, $\mathrm{Pure}_{\mathrm{SNIa}}$, $\mathrm{Pure}_{\mathrm{NSM}}$, and $\mathrm{Pure}_{\mathrm{oldAGB}}$, defined by $f_k(\mathrm{Pure}_j)=\delta_{kj}$ corresponding to the single fluids described in \autoref{ssec:methods}. These synthetic elements are helpful to include because they provide a longer baseline in $\boldsymbol{\Delta}$, thereby improving fit quality. This is especially important for channels that do not provide a majority contribution to more than one or two of the elements we follow (e.g., SNIa and NSM). For these pure elements, we compute $\boldsymbol{\Delta}$ and measure the stellar and gas correlations in the simulations exactly as we do for the other nine elements.

\begin{figure}
    \centering
    \includegraphics[width=1\columnwidth]{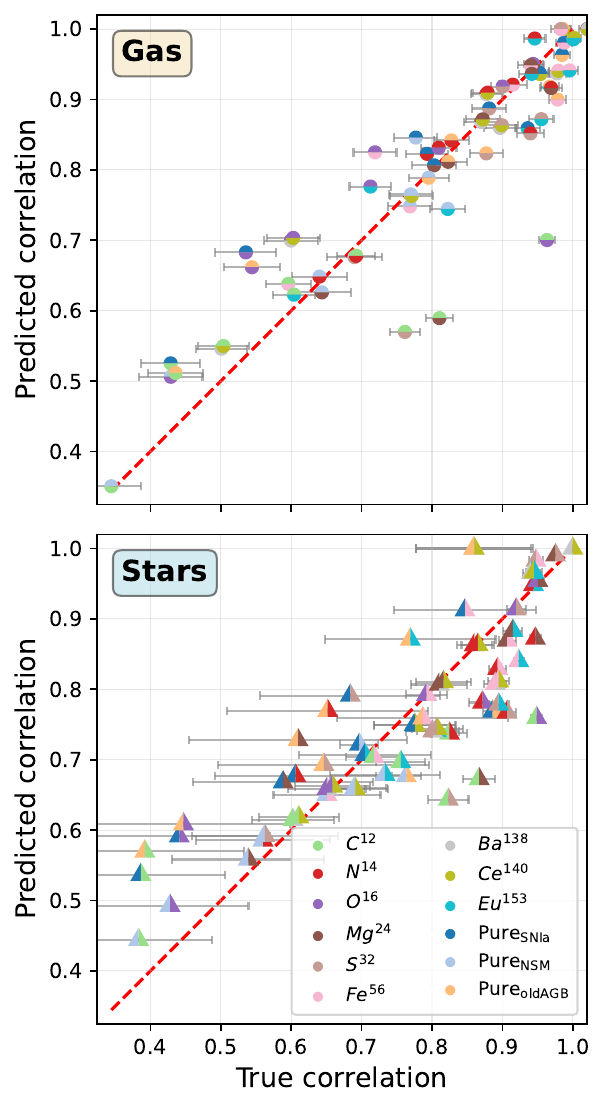}
    \caption{These panels show the true element-element correlations measured from the simulations on their horizontal axes in the gas phase (top) and stars (bottom), while the vertical axes of these panels show the correlation predicted by our simple linear regression model (\autoref{eq:linear_regression}). Each point corresponds to a particular element pair, with the colours in the half-circle or half-triangle markers indicating the elements being compared (see legend). The horizontal coordinates of the points represent the time average over the period from $501$--600 Myr in the simulations, and the horizontal error bars show the standard deviation over this time. The red dashed line is the one-to-one line for comparison.}
    \label{fig:linear_regress}
\end{figure}

We now fit for $\mathbf{c}$ and $b$ over all element pairs (including our newly-introduced ``pure'' elements) and all times, using as input data to the fit the measured cross-correlations between each element pair at the same times for which we have measured $\boldsymbol{\Delta}$; we carry out separate fits for gas and stars. We estimate central values for $\mathbf{c}$ and $b$ using ordinary least squares fitting, while to estimate uncertainties we use a bootstrapping method: for each iteration, we independently sample $N$ elements pairs with replacement from the full dataset (where $N$ is the total number of pairs), then fit the constrained regression model \autoref{eq:linear_regression} using a minimum-loss procedure with bounds $c_k \in (-\infty, 0]$. From the resulting 200 coefficient vectors, we extract lower and upper confidence bounds using the 16th and 84th percentiles, corresponding to $\pm 1\sigma$ in a normal distribution. We report the fit coefficients and uncertainties obtained from this procedure in \autoref{tab:linear_regress_coeff}, and compare the predictions made by our best-fitting models for gas and stars in \autoref{fig:linear_regress}; the model predictions have been clipped to the range $[0,1]$.

This exercise allows us to make a few observations. First, despite its simplicity, the model is remarkably good at predicting how well different element pairs correlate: the $R^2$ values that characterise the quality of fit are 0.84 and 0.73 for gas-phase and stellar element-element correlations, respectively. Second, the fact that the fit coefficients are all negative encodes the fundamental principle that larger differences in nucleosynthetic origin suppress element correlation: element pairs with small $\boldsymbol{\Delta}$ yield predictions closer to unity (high correlation), while pairs with large compositional divergence yield smaller correlations. Third, the relative sizes of the coefficients of $\mathbf{c}$ reflects the relative importance of each channel: WR shows the strongest discriminative power (largest $|c_k|$), followed by SNII and NSM, while SNIa contributes weakly. This hierarchy arises because WR and SNII dominate carbon and $\alpha$ elements respectively, which have very distinct spatial and temporal signatures, followed by rare NSM merger events. By contrast both SNIa and AGB (both old and young) channels have smaller effects on element-element correlation, likely due to the smaller spatial and temporal variations they induce in the distributions of the elements they produce.

Notably, cross-correlations between C and elements produced by SNII are the most significant outliers in both the gas and stellar fits; for these, the model systematically over-predicts the cross-correlation, with a larger bias in the gas phase. No comparable bias is seen for cross-correlations between C and elements not dominated by SNII production. We therefore attribute the failures of the model for C-SNII correlations to the extremely short injection intervals between WR and SNII. Carbon injected by WR winds has little time to establish an independent spatial pattern before nearby SNII events occur. The subsequent SNII feedback then dynamically re-arranges and effectively overwrites the recent WR imprint, so the carbon field is governed by stochastic SNII-driven transport rather than by the smoother co-production picture assumed by the model. In a gas-phase two-point statistic, this dynamical overwriting can make the measured C-SNII cross-correlation exceed the model prediction, yielding the observed negative residual; the same tendency is present but weaker for stellar abundances, a phenomenon that we show in the next section is part of a general trend for differences between stellar and gas-phase cross-correlations among elements injected in SNII explosions.

\begin{table}
    \centering
    \caption{Best-fit linear regression coefficients $c_k$ for predicting gas-phase and stellar element-pair cross-correlations from total mass-weighted source-fraction differences. Quoted uncertainties are derived from bootstrapping as discussed in the main text. More negative $c_k$ values imply stronger decorrelation for greater compositional divergence.}
    \label{tab:linear_regress_coeff}
    \begin{tabular}{llr}
        \hline
        Nucleosynthetic Channel & \multicolumn{2}{c}{Decorrelation Coefficient} \\
        & $c_{k,\mathrm{gas}}$ & $c_{k,\mathrm{star}}$ \\
        \hline
        WR    & $-0.77^{+0.07}_{-0.07}$ & $-0.53^{+0.10}_{-0.08}$ \\
        SNII  & $-0.36^{+0.07}_{-0.06}$ & $-0.40^{+0.09}_{-0.06}$ \\
        AGB   & $-0.16^{+0.07}_{-0.06}$ & $-0.25^{+0.11}_{-0.10}$ \\
        oldAGB& $-0.11^{+0.03}_{-0.03}$ & $-0.19^{+0.05}_{-0.04}$ \\
        SNIa  & $-0.05^{+0.03}_{-0.02}$ & $-0.15^{+0.04}_{-0.03}$ \\
        NSM   & $-0.23^{+0.02}_{-0.02}$ & $-0.24^{+0.03}_{-0.02}$ \\
        \hline
    \end{tabular}
\end{table}

\subsection{Differences between elemental cross-correlations in gas and stars}

The comparison between gas-phase and stellar-phase model coefficients reveals that, while the linear regression model achieves similarly good fits for both phases ($R^2 = 0.84$ for gas versus $0.73$ for stars), the fit coefficients themselves differ substantially, as do the values of the cross-correlations themselves. This suggests that although source-fraction differences drive correlations in both gas and stars, additional physical processes modulate how these source-compositional differences manifest. The sensitivity test in \aref{app:oldagb_sensitivity} shows that this conclusion is not specific to the adopted old-star AGB return rate: across $f=0$--2 the model retains $R^2=0.70$--0.88 in gas and $0.71$--0.77 in stars.

\begin{figure}
    \centering
    \includegraphics[width=1\columnwidth]{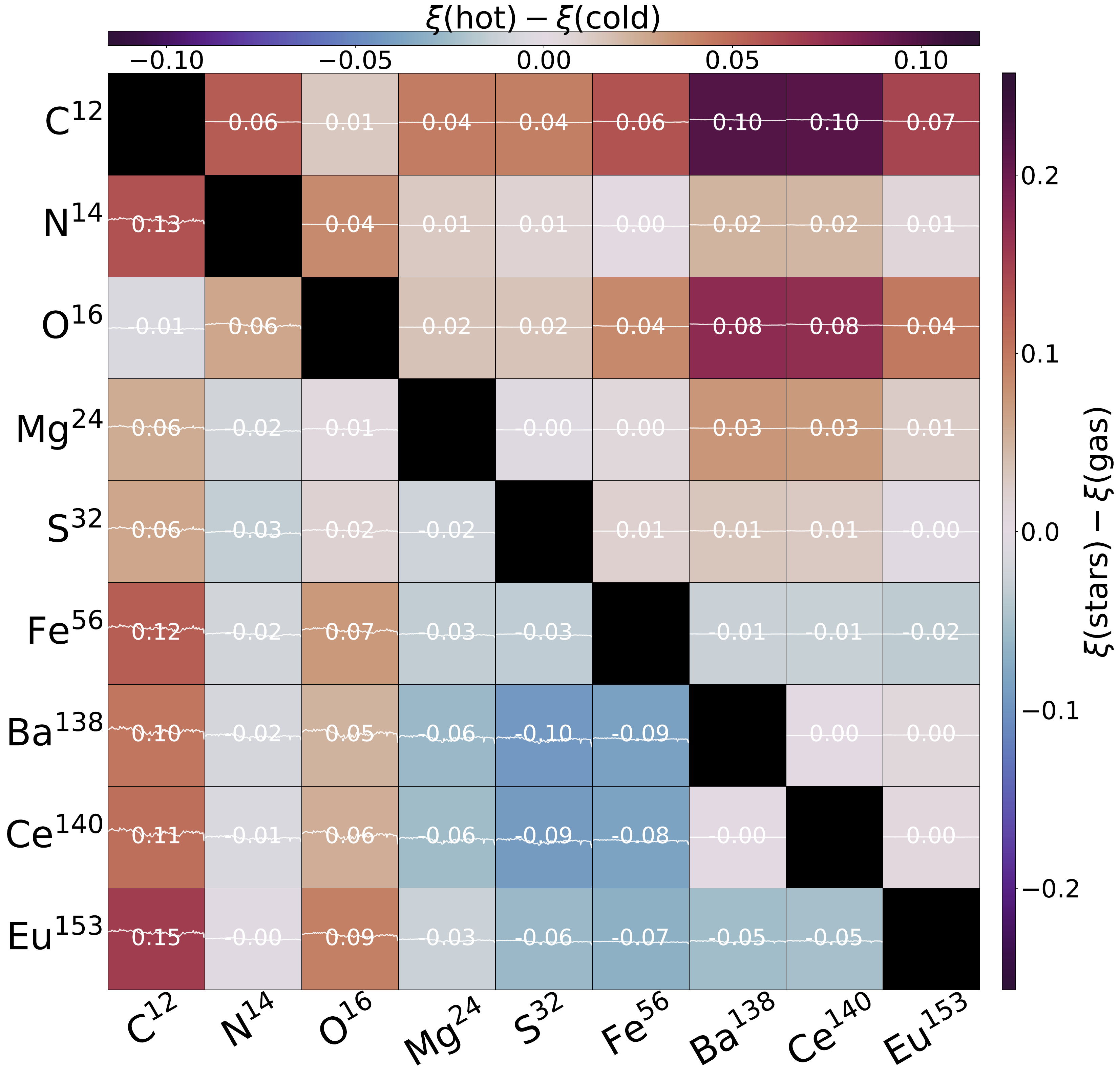}
    \caption{To visually divide this table into two mirroring halves, the diagonal row has been left blank. The lower-left triangle shows the difference between the stellar and gas-phase cross-correlations, $\xi_{\mathrm{star}}-\xi_{\mathrm{gas}}$ for each element pair; the white line in each cell shows the time evolution of the difference, with the horizontal axis running from $501$--$600$ Myr, and the background colour and numerical value give the mean value over the final 100 Myr. The upper-right triangle of cells are identical, except that rather than the difference between stellar and gas-phase cross-correlations show the difference between the gas-phase correlations for hot ($T>1000$ K) and cold ($T<1000$ K) gas, $\xi_{\mathrm{hot}}-\xi_{\mathrm{cold}}$ -- see main text for details. Note that the star-gas and hot-cold differences use different colour scales -- the horizontal colour bar is for hot-cold, and the vertical one for star-gas -- but that both colourmaps are symmetric about zero.
    }

    \label{fig:stellar_minus_gas_correlation_diff}
\end{figure}

To investigate the origin of differences between gas and stellar correlations, we plot the differences for each element pair in the lower left part of \autoref{fig:stellar_minus_gas_correlation_diff}.
This figure reveals a striking pattern: C and O are systematically more highly correlated with other elements in the stellar phase than in the gas phase, while Fe and Eu show the opposite trend, with little difference between stellar and gas correlation for most elements; intermediate elements exhibit characteristics grouped by nucleosynthetic origins. This element-dependent divergence between phases is not a random effect but reflects fundamental differences in how rapidly different enrichment channels inject materials and how quickly the ISM mixes them.

The tendency for C and O to be more correlated in stars than in gas arises from their rapid, early injection. C is dominated by WR winds and O by SNII, both of which operate on short timescales (few Myr to tens of Myr following star formation). These rapid injection events establish strong spatial correlations in the ISM at early times. However, when these enriched materials are injected, they often go into large-scale supernova-driven bubbles that are initially hot ($T \gtrsim 10^6$ K) and expand across the disc. These hot, bubble-filled regions suppress the gas-phase correlation with other elements because the hot ISM is less well-mixed than the cold ISM. However, since these hot bubbles do not efficiently form stars, there is no comparable suppression in the stellar phase: stars forming in the islands of dense gas between bubbles inherit the chemical correlations established present in colder gas that has had more time to mix. Thus $\langle \xi_\mathrm{star}\rangle > \langle \xi_\mathrm{gas}\rangle$ for C and O. This interpretation is also consistent with the scenario proposed by \citet{Metha2026}, in which metal-rich hot bubbles produced by core-collapse supernovae dominate short-timescale metallicity fluctuations.

Conversely, Fe and Eu are delayed-source elements dominated by SNIa and old AGB stars. These channels are distributed in both space and time, injecting material continuously over the full simulation and throughout the volume of the gas, regardless of whether it is cold and star-forming or hot and non-star-forming. Because they are not associated with the large-scale gaseous structures created by star formation feedback, they show relatively little difference between their correlations in gas and stars. Elements injected by younger AGB stars fall between these two limiting cases, with some mild association with structures established by star formation feedback.

To check the validity of this interpretation, we partition the gas into hot and cold components using a temperature threshold of 1000 K and compute Pearson correlations separately for each phase. Unlike the gas-phase correlations we have discussed thus far, which are computed on projections and thus directly comparable to observations, for this purpose we instead compute the elemental correlations exactly as we do for stars in \autoref{ssec:analysis_crosscor}, i.e., we collect the set of gas particles present at every time snapshot, divide them into hot and cold phases, and then for each phase we compute the cross-correlation between each element pair for the gas particles using \autoref{eq:crosscorr_stars}. This quantity is not observable, but it is informative of how the correlation structure of elements differs between hot and cold gas. We show the result in the upper-right triangle of \autoref{fig:stellar_minus_gas_correlation_diff}). The figure shows that the pattern of differences between hot and cold gas correlations (upper-right triangle) mirrors the pattern of differences between stellar and total gas correlations (lower-left triangle) with striking fidelity. This correspondence confirms that the star-gas divergence is fundamentally a thermal effect: stars preferentially form in the cold ISM phase and thus inherit the chemical correlations present in cold gas, while the hot ISM contributes to the gas-phase spatial correlation but does not form stars. As a result, the elements that are more strongly correlated in stars than in gas are precisely those that are more strongly correlated in hot gas than in cold.

The cross-correlation evolution further supports this picture. Gas-particle internal correlations and stellar correlations converge quickly and remain comparatively stable, whereas gas two-point spatial correlations (see in \autoref{fig:cross_corr_gas_2d}) show much larger temporal fluctuations. This difference indicates that two-point statistics are especially sensitive to large-scale transport and structural reconfiguration driven by feedback (such as the hot bubbles). Consistently, element pairs with larger hot--cold phase contrast also show stronger time variability in gas two-point cross-correlation, linking thermal segregation directly to the amplitude of temporal decorrelation.

\subsection{Comparison with Observations}
\label{ssec:sim_vs_obs}

Our final discussion topic is a comparison of the stellar cross-correlations found in our simulations with those observed in Milky Way stars. For our observational comparison data, we use two data sets. The first is a collection of 560 Solar-type stars from \citet{Casali2020}, who derive abundances and isochronal ages from high-resolution HARPS spectra \citep{Mayor2003} retrieved from the European Southern Observatory (ESO) archive; the elements that overlap between their data and our simulations are C, Mg, S, Fe, Ba, Ce, and Eu. The second is a selection 199,289 red giant stars with abundances for 16 elements from \citet{GG2026}, which are selected using quality cuts on signal to noise, stellar parameters, flags and abundance uncertainty, from the SDSS V Galactic Genesis survey \citep{DR19, Meszaros2025}; the elements in common between this sample and our simulations are O, Mg, S, Fe, and Ce. The ages for each of the stars are from \citet{StoneM2025}. Since carbon and nitrogen are the primary age-dating features in that work, we exclude C and N from this sample to avoid circularity. For both sets of data, we group the stars into age bins 1 Gyr wide and calculate the cross-correlations in each age bin using \autoref{eq:crosscorr_stars}. We then take the average cross-correlation over all age bins as our final observational estimates. Because the datasets lack stellar mass measurements, we assign equal weight to every simulated star particle and every observed star for this comparison, rather than using the mass-weighted simulated statistics adopted elsewhere, ensuring that both axes correspond to the same object-weighted sampling.

\begin{figure}
    \centering
    \includegraphics[width=1\columnwidth]{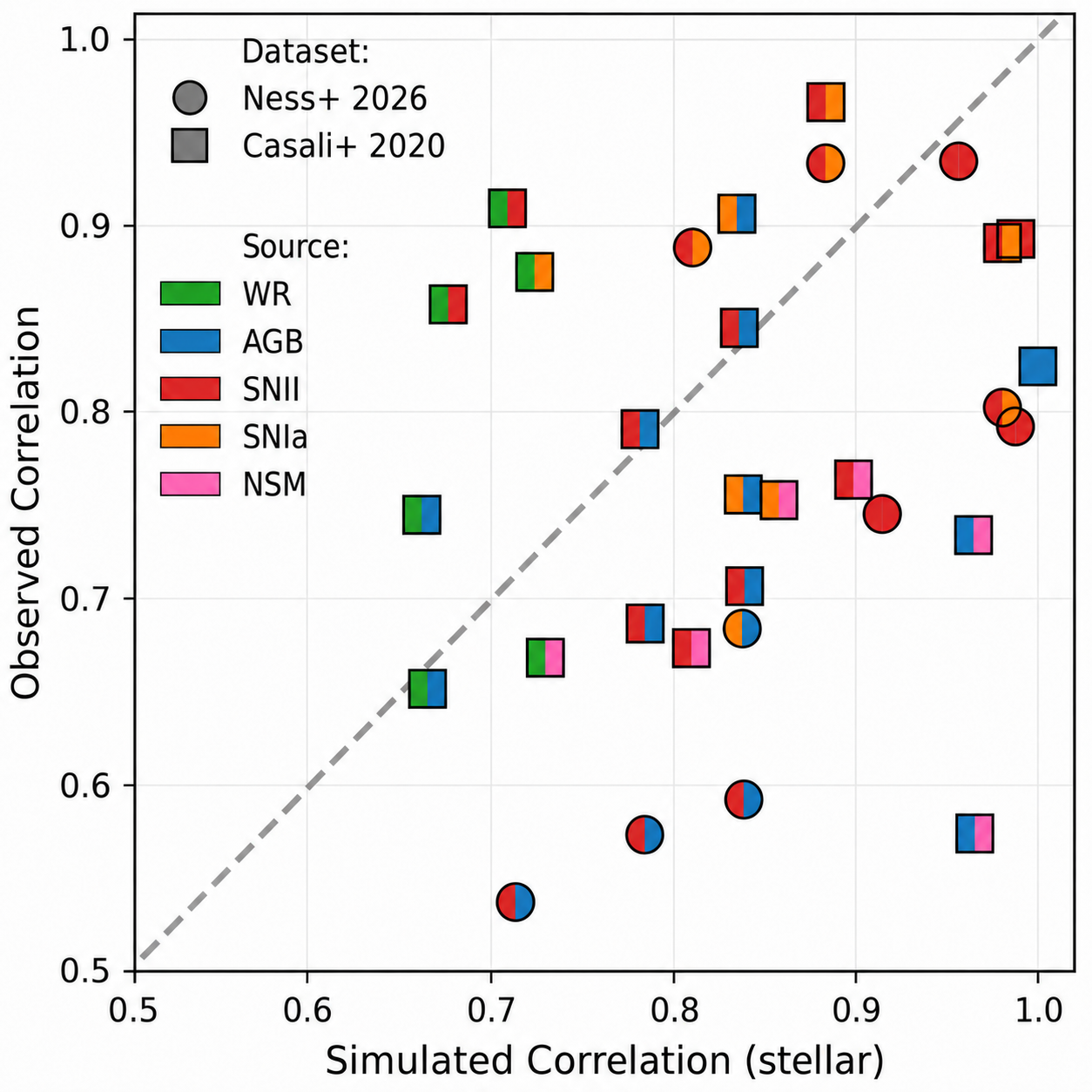}
    \caption{Comparison of simulated versus observed element-pair Pearson correlations. Each point plots the simulated stellar correlation (x-axis) against the observed correlation (y-axis) for the same element pair. The dashed diagonal is the 1:1 reference line, and departures from this line indicate mismatches in pairwise chemical correlation structure between the simulations and the observations. Marker shape identifies the observational dataset: circles for \citet{GG2026}, squares for \citet{Casali2020}. The individual elements plotted are O, Mg, S, Fe, Ce for the \citeauthor{GG2026} data set and C, Mg, S, Fe, Ba, Ce, Eu for the \citeauthor{Casali2020} one. Each marker is split into two colours to indicate the dominant nucleosynthetic channels producing each of the two elements being compared (see legend): WR stars for C, SNII for O, Mg, and S, AGB stars for N, Ba, and Ce, SNIa for Fe, and NSM for Eu. (Note that, although \autoref{fig:source_fractions} shows that NSM does not dominate over AGB for Eu, we nonetheless colour it as NSM because it is the only element in our sample for which NSM makes a significant contribution.)
    }
    \label{fig:sim_obs_compare}
\end{figure}

We compare the simulated and observed element-pair correlations in \autoref{fig:sim_obs_compare}. Before even considering the comparison to simulations, it is important to note that the two observational datasets differ substantially from each other, even for the same element pair. For those element pairs in common between the two samples, the typical inter-dataset scatter in correlation coefficient is $\sim 0.05$--$0.15$, with correlations from \citeauthor{Casali2020} generally higher than from \citeauthor{GG2026}. These discrepancies are comparable to, and often exceed, the simulation-observation offsets visible in the figure. This inter-dataset scatter reflects the combined effects of differing spectral resolution and wavelength coverage, distinct stellar populations and evolutionary stages being probed, heterogeneous stellar parameter pipelines, and different selection functions across the Galactic disc. Consequently, it is not immediately clear whether apparent tensions between simulation and observation reflect genuine deficiencies in our model or instead arise from systematic uncertainties inherent in the observational abundance measurements themselves. We explore these systematics and their possible origins in more detail in \aref{app:obs_discrepancies}.

With this caveat in mind, the overall level of agreement between simulations and observations is encouraging, and indicates that RoSE has successfully captured the primary processes influencing the chemical distribution structure of galaxies. In particular, in both the simulations and observations we generally find that correlations are highest between elements from the same nucleosynthetic group (i.e., solid-coloured points lie preferentially at the top-right of the plot), correlations between SNII and SNIa elements are on average the next-highest, and correlations between SNII and AGB elements are lowest.

The largest systematic tensions between the simulations and observations are for SNII-AGB pairs from the \citeauthor{GG2026} sample (blue and red circles), and for pairs involving WR or NSM elements from the \citeauthor{Casali2020} data set (squares including pink or green). For the SNII-AGB pairs, the simulation tends to predict somewhat stronger correlations than the \citet{GG2026} observations, with points falling below the one-to-one line. There is significantly better agreement with the \citeauthor{Casali2020}~results for SNII-AGB pairs, highlighting the importance of systematic disagreements within the observations. For the pairs including WR or NSM elements, when compared with the \citeauthor{Casali2020}~dataset RoSE appears to underestimate the correlation for WR element pairs while overestimating that of NSM pairs. Several physical and observational mechanisms could contribute to these offsets. On the observational side, AGB-sourced elements (Ba, Ce) are typically measured from weaker, more blended spectral features, and their formal abundance uncertainties are larger than for the SNII (O, Mg, S); measurement noise artificially suppresses observed correlations, preferentially affecting these mixed-source pairs. Additionally, the two observational samples probe different radial and age distributions within the Milky Way disc: the SDSS-V red giants span a broader range of Galactic radii and ages than the Solar-neighbourhood, Solar-type stars of \citeauthor{Casali2020}, introducing radial abundance gradients and age-metallicity scatter that further dilute pairwise correlations. On the simulation side, the RoSE model tracks only nine isotopes across six nucleosynthetic channels, potentially underestimating the decorrelation produced by the full chemical dimensionality of real stellar populations; and the isolated, relatively quiescent galactic environment of RoSE over 600 Myr does not capture bar-driven radial migration or satellite-driven kinematic heating, both of which would admix chemically distinct populations and weaken correlations in observed samples. 

Taken together, however, the comparison between simulations and observations shown in \autoref{fig:sim_obs_compare} demonstrates that the nucleosynthetic source framework established in this work, wherein same-channel pairs are strongly correlated and mixed-channel pairs are weaker, is qualitatively validated by current observational data. The residual quantitative offset in mixed-source pairs, while genuine, is no larger than the inter-dataset scatter between the two independent observational samples, and thus cannot presently be attributed unambiguously to shortcomings in RoSE. Future surveys delivering high-fidelity multi-element abundance measurements, together with precise stellar ages and sample sizes large enough to support statistical analyses such as those presented here, will be essential for testing whether these mixed-source tensions reflect physical processes missing from RoSE or rather the significant systematic challenges of extracting robust correlation measurements from observed stellar spectra.

\subsection{Scope and caveats}
\label{ssec:caveats}

We close the discussion by making explicit the intended scope and limitations of RoSE. Our simulation is intended to be a controlled, high-resolution experiment on an isolated Milky Way-mass disc, and while the initial condition is designed to reproduce the bulk characteristics of the Milky Way, it is not tuned to reproduce its kinematic structure or assembly history in detail. For example, our simulation does not include a bar, perturbations from satellite galaxies, or ongoing cosmological gas accretion or mergers. These missing processes are important for determining the present-day distribution of old stars in both chemical and physical space: bar-driven radial migration, satellite heating, and mergers all mix stellar populations born at different radii and epochs, thereby changing quantitative correlations (for example between abundances, age, and velocity dispersion) measured in a Galactic stellar sample that is selected over a fixed volume rather than at fixed stellar age. Moreover, even if we did include these processes, it is unclear that the 600 Myr duration of our simulation would be sufficient to capture their effects. It is for this reason that we have not attempted to examine the statistical properties of mixed-age stellar populations in this study, and have focused instead only on the gas and the abundances of mono-age stellar populations. Cosmological simulations specifically tuned to match the Milky Way \citep[e.g.,][]{2017MNRAS.467..179G, Agertz21a} include these assembly and environmental effects and are generally better suited to such questions, making them complementary to the controlled experiment studied here.

Conversely, the spatial and mass resolution practical in current cosmological campaigns generally does not permit the same star-by-star treatment of element return or the small-scale gas and young-star correlation statistics that are our principal focus here. The goal of this paper is to measure how source-dependent enrichment and turbulent mixing structure gas and newly formed stars once the simulated disc reaches a quasi-steady injection-mixing state. This does not require simulation durations approaching the Hubble time -- \autoref{fig:source_fractions}--\autoref{fig:stellar_cross}, together with the temporal-correlation analysis in \autoref{ssec:results_temporal}, show that the relevant source fractions, spatial correlation parameters, gas cross-correlations, and stellar cross-correlations settle into a quasi-steady state within a few hundred Myr. The resolution of RoSE allows individual supernovae, star-forming giant molecular clouds, and the multiphase ISM to be represented more explicitly than is currently practical in cosmological Milky Way simulations. RoSE should therefore be viewed as occupying one end of a hierarchy of complementary approaches: controlled isolated simulations resolve local enrichment and mixing, while cosmological campaigns such as Auriga \citep{2017MNRAS.467..179G} and Vintergatan \citep{Agertz21a} capture assembly history and environment.

Another important caveat regards our treatment of nucleosynthesis and element injection. Our set of nine isotopes and five nucleosynthetic channels is chosen to provide good coverage of the main enrichment channels for most elements, but it does omit a number of other nucleosynthetic pathways, such as classical novae, cosmic-ray spallation, and hypernovae, which may be dominant for certain isotopes. Moreover, we have explored only a single relatively simple set of yield tables, and have not explored either how the results would change for different yield tables, or for more complex yield models that include, for example, multiple SNIa subclasses. These omissions are most important for elements not tracked here, for detailed isotopic ratios, and for the interpretation of elements whose real production may involve multiple rare channels. Our treatment of NSMs is also necessarily very approximate, and relies on the approximation of a boosted rate in order to improve sampling of the spatial distribution from this, rare delayed channel. Results based on NSM production should therefore be treated as tentative.

Finally, RoSE does not model the formation, destruction, or transport of dust and therefore does not distinguish gas-phase oxygen from oxygen locked in grains. By contrast, extragalactic nebular abundance maps measure only gas-phase oxygen. Estimates for H\,{\sc ii} regions imply oxygen depletions of approximately $0.08$--$0.12$ dex, with the depletion tending to increase with metallicity \citep{Peimbert2010Dust}; grain growth and destruction can also make depletion depend on local ISM conditions. Uniform depletion would have little effect on the fluctuation statistics used here, but spatially varying depletion could alter the measured oxygen autocorrelation, correlation length, and injection scale. Comparisons between RoSE's total-oxygen structure and observed gas-phase oxygen maps should therefore not be interpreted as exact one-to-one tests.

\section{Conclusions}

On the canvas of a galactic disc, stellar enrichment leaves measurable, rhythmically patterned imprints: stellar winds fall like fine drizzle, slowly and locally moistening their surroundings, while supernovae burst like sudden storms, instantaneously driving strong dynamical perturbations and wide-ranging mass transport; short-lived injection events behave like transient showers, whereas delayed channels resemble prolonged rainy seasons that accumulate and reshape the chemical landscape over long intervals. Different elements, governed by their nucleosynthetic channels, paint distinguishable spatial ``ripples'' across the disc -- ripples that encode both source information and the timescales of mixing and transport. RoSE (Ripples of Stellar Enrichment) follows this process, revealing how nucleosynthetic channels, phase segregation, and dynamics jointly determine inter-element correlations and their persistence.

This study demonstrates that the spatial, temporal, and cross-correlation structure of elemental fluctuation distributions in galaxies is fundamentally organized by the diversity of nucleosynthetic sources. 
Our core findings are threefold:
\begin{enumerate}
\item Nucleosynthetic source clustering is the primary organizing principle of elemental spatial structure. Elements produced by the same nucleosynthetic channel (e.g., SNII-dominated O, Mg, S or AGB-dominated Ba, Ce) cluster tightly in spatial statistics, with auto-correlation parameters closely tracking those of the dominant source. Individual nucleosynthetic channels imprint characteristic length scales and timescales into the ISM that persist long enough for the elemental spatial distribution to act as a fossil record of source-specific injection environments.
\item The differences or similarities between elements' nucleosynthetic origin similarly determine the cross-correlation between their abundances in both the interstellar medium (ISM) and young stars. A simple linear regression model relating the cross-correlation between element pairs to the level of similarity or difference in their nucleosynthetic origins predicts those cross-correlations remarkably well ($R^2 = 0.84$ in gas, $0.73$ in coeval stars). The predictive power of this model is fundamentally unchanged between gas and stellar phases. Varying the assumed old-star AGB return rate between zero and twice its fiducial value changes individual correlations but preserves the element-pair ordering and the explanatory power of the source-fraction model. This result aligns qualitatively with the analytic \citet{KT25} model wherein the differences in delay time between star formation and element return for different astrophysical origin sites governs the spatial and correlation structure of elemental abundances.
\item There are, however, noticeably differences between the abundance correlations found in stars and the interstellar medium, with most element pairs showing higher correlation in stars than in ISM gas. We show that this difference arises from the temperature-dependent environment of star formation. Supernova feedback produces large-scale hot bubbles of enriched gas, which lower the correlation in the gas phase; however, these hot phases do not form stars, so young stellar populations form in cooler, more-mixed gas and inherit stronger correlations. The extent to which elements are better correlated in stars than in gas is therefore primarily a function of whether the nucleosynthetic channels producing those elements also produce large bubbles of very hot gas.
\end{enumerate}

We also compare our simulations results with observations of coeval Milky Way stars. This comparison qualitatively validates the nucleosynthetic source framework, while the remaining quantitative discrepancies are modest, comparable to the level of disagreement between independent observational data sets, and plausibly explained by uncertainties in stellar age estimation in the observational samples.
Consequently, current observations do not yet allow us to unambiguously distinguish whether these residual differences reflect physical processes missing from RoSE (such as bar-driven radial migration) or instead arise from abundance and age systematics in the observational data, underscoring the need for high-fidelity multi-element abundances and precise ages.

Looking forward, the success of the nucleosynthetic source framework motivates several promising research directions. In simulations, natural extensions include: (1) cosmological zoom-in simulations at comparable resolution to test whether our results depend critically on galactic environment and feedback intensity; (2) inclusion of additional physics such as bar-driven radial migration and AGN feedback to systematically isolate which processes account for any simulation-observation residuals; (3) extension to longer timescales (multi-Gyr) to examine the asymptotic mixing and correlation behaviour achieved over the lifetimes of observed stellar populations; and (4) parametric studies varying feedback strength, star-formation efficiency, and subgrid turbulence to understand how nucleosynthetic imprints depend on galactic context. Observationally, the framework makes firm predictions that can be tested by upcoming instrumentation: high-resolution integral-field unit surveys (e.g., BlueMUSE; \citealt{BlueMUSE}) can map multiple elements across nearby galaxies with minimal selection bias; and next-generation stellar abundance surveys (e.g., 4MOST; \citealt{4MOST2019}, WEAVE-StePS; \citealt{Iovino2023}, SDSS-V; \citealt{Kollmeier2026}) can provide homogeneous multi-element abundances and precise ages for tens of thousands of Galactic stars with well-defined selection functions. These observations will enable robust tests of whether the nucleosynthetic source framework extends to real galactic environments and clarify the precise roles of thermal physics, dynamics, radial migration, and small-scale turbulent transport in shaping galactic chemical structure. The RoSE project demonstrates that high-resolution simulations tracking star-by-star nucleosynthetic feedback provide an essential bridge between the predictive power of analytic source-delay theory and the complexity of real observations, enabling us to decode the enrichment histories written in the spatial and chemical structure of galaxies.

\section*{Acknowledgements}

We acknowledge high-performance computing resources provided by the Australian National Computational Infrastructure (award jh2) through the National and ANU Computational Merit Allocation Schemes and Astronomy Supercomputer Time Allocation Committee (ASTAC) in Australia. Some of the text and code refinements were developed in micro-managed highly supervised engagement with AI language models (Claude, Anthropic and Gemini, Google), but all final text was generated directly or edited heavily by the human authors.

CZ acknowledges financial support from the Research School of Astronomy and Astrophysics supplementary scholarship, Australian National University postgraduate research scholarship and Chinese Scholarship Council. ZL acknowledges the Science and Technology Facilities Council (STFC) consolidated grant ST/X001075/1. MRK acknowledges support from the Australian Research Council through Laureate Fellowship FL220100020. ZH acknowledges support from National Natural Science Foundation of China (NSFC) through grant No. 12503026 and support from Boya Fellowship at Peking University. This research was supported by NSF Grant AST-2406729 and a Humboldt Research Award from the Alexander von Humboldt Foundation.

Thanks to the observations collected with the FLAMES instrument at VLT/UT2 (Paranal Observatory, ESO, Chile), for the Gaia-ESO Large Public Spectroscopic Survey (188.B-3002, 193.B-0936).

Funding for the Sloan Digital Sky Survey V has been provided by the Alfred P. Sloan Foundation, the Heising-Simons Foundation, the National Science Foundation, and the Participating Institutions. SDSS acknowledges support and resources from the Center for High-Performance Computing at the University of Utah. SDSS telescopes are located at Apache Point Observatory, funded by the Astrophysical Research Consortium and operated by New Mexico State University, and at Las Campanas Observatory, operated by the Carnegie Institution for Science. The SDSS web site is \url{www.sdss.org}.

SDSS is managed by the Astrophysical Research Consortium for the Participating Institutions of the SDSS Collaboration, including the Carnegie Institution for Science, Chilean National Time Allocation Committee (CNTAC) ratified researchers, Caltech, the Gotham Participation Group, Harvard University, Heidelberg University, The Flatiron Institute, The Johns Hopkins University, L'Ecole polytechnique f\'{e}d\'{e}rale de Lausanne (EPFL), Leibniz-Institut f\"{u}r Astrophysik Potsdam (AIP), Max-Planck-Institut f\"{u}r Astronomie (MPIA Heidelberg), Max-Planck-Institut f\"{u}r Extraterrestrische Physik (MPE), Nanjing University, National Astronomical Observatories of China (NAOC), New Mexico State University, The Ohio State University, Pennsylvania State University, Smithsonian Astrophysical Observatory, Space Telescope Science Institute (STScI), the Stellar Astrophysics Participation Group, Universidad Nacional Aut\'{o}noma de M\'{e}xico, University of Arizona, University of Colorado Boulder, University of Illinois at Urbana-Champaign, University of Toronto, University of Utah, University of Virginia, Yale University, and Yunnan University.
\section*{Data Availability}

The data underlying this article will be shared upon reasonable request to the corresponding author.



\bibliographystyle{mnras}
\bibliography{example} 




\appendix

\section{Sensitivity to the old-star AGB return rate}
\label{app:oldagb_sensitivity}

The prescription in \autoref{ssec:methods} assigns old star particles a time-averaged AGB element return rate per unit stellar mass. To quantify the impact of this assumption, we introduce a multiplier $f$ applied to the old-star AGB contribution and repeat the elemental cross-correlation analysis for $f=0$, $0.5$, $1$, and $2$, where $f=1$ is the fiducial model used throughout the main text. This is a post-processing sensitivity test: for every value of $f$ we use the same source-resolved simulation outputs and multiply the old-star AGB scalar contribution by $f$ when reconstructing total element abundances and source fractions. The test therefore isolates the dependence of the reported chemical statistics on the adopted return rate; it does not rerun the dynamics, which are unchanged because the tracked metal fields are passive scalars.

For the gas we reconstruct 20 kpc $\times$ 20 kpc abundance maps and recompute the zero-lag cross-correlations for snapshots $t=501$--600 Myr using the procedure in \autoref{ssec:analysis_crosscor}. For stars we use the $t=600$ Myr snapshot, divide stars of age 0--100 Myr into 1 Myr cohorts, and compute the mass-weighted correlation defined by \autoref{eq:crosscorr_stars}. Thus both phases sample the same birth epochs. We also repeat the regression analysis of \autoref{eq:linear_regression}, including the three pure basis elements $\mathrm{Pure}_{\mathrm{SNIa}}$, $\mathrm{Pure}_{\mathrm{NSM}}$, and $\mathrm{Pure}_{\mathrm{oldAGB}}$ whenever their corresponding channels are present. For $f=0$, the old-star AGB contribution vanishes identically, so we remove both $\mathrm{Pure}_{\mathrm{oldAGB}}$ and the oldAGB predictor from the regression; this fit therefore contains 11 variables, five source channels, and 55 unique pairs, compared to 12 variables, six channels, and 66 pairs for $f>0$.

\begin{figure}
    \centering
    \includegraphics[width=1\columnwidth]{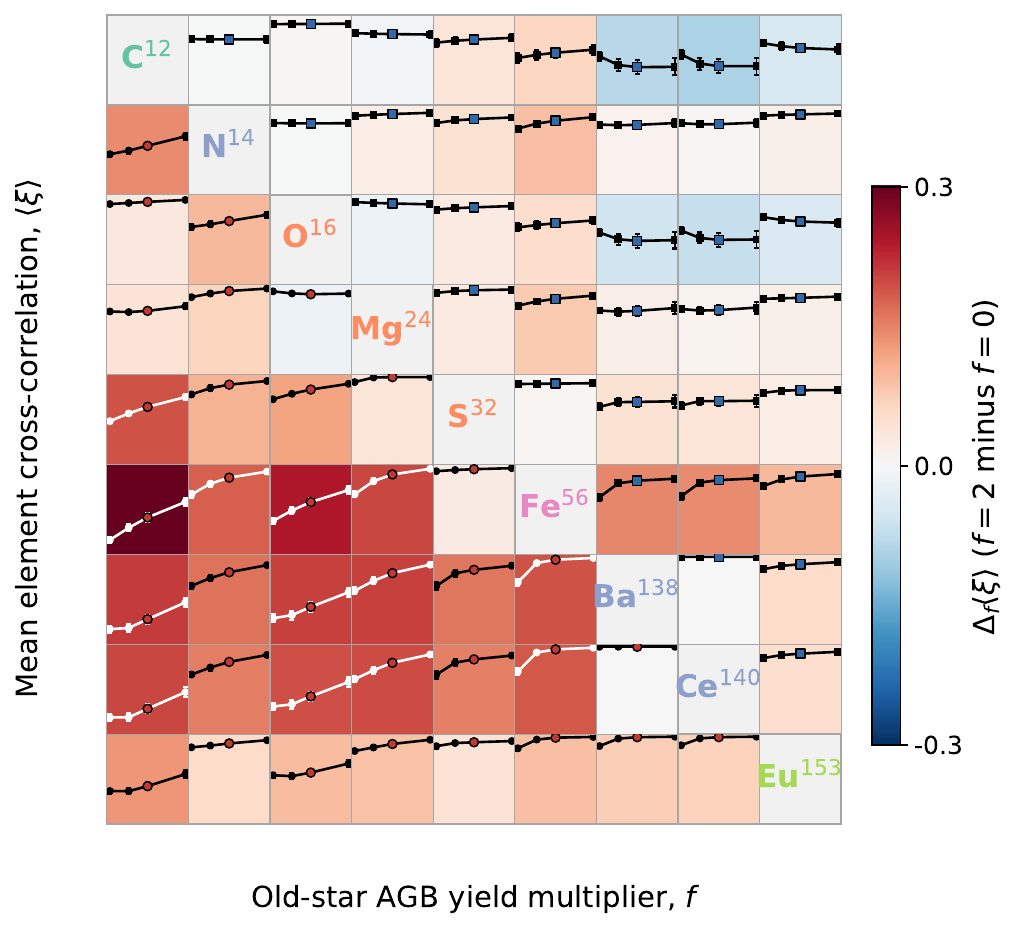}
    \caption{Sensitivity of the element-element cross-correlations to the old-star AGB yield multiplier $f$. Each off-diagonal panel shows the mean correlation over the final 100 Myr as a function of $f$, with error bars giving the standard deviation among the 100 individual snapshots (gas; circles in the lower triangle) or 1 Myr stellar age cohorts (mass-weighted stars; squares in the upper triangle). Black-edged points mark the fiducial $f=1$ model. The background colour encodes $\Delta_f\langle\xi\rangle=\langle\xi\rangle_{f=2}-\langle\xi\rangle_{f=0}$ on a zero-centred symmetric scale: red denotes increasing correlation with $f$, while blue denotes decreasing correlation.}
    \label{fig:oldagb_sensitivity_matrix}
\end{figure}

\autoref{fig:oldagb_sensitivity_matrix} shows that the response is smooth but element-pair dependent. Across the full range $f=0$--2, the largest gas-phase excursion is $0.305$ for C--Fe, while the median excursion over all 36 element pairs is $0.140$. The stellar correlations are less sensitive: their largest excursion is $0.149$ for Fe--Ba and their median excursion is $0.039$. The largest changes generally involve pairs whose source fractions receive appreciable old-star AGB contributions, whereas pairs dominated by a shared prompt source remain close to unity. Importantly, pairs assigned to the same dominant nucleosynthetic channel remain more strongly correlated on average than mixed-source pairs for every tested value of $f$.

\begin{table*}
    \centering
    \caption{Summary of the old-star AGB sensitivity test. Correlation changes and Spearman rank coefficients compare the 36 real-element pairs at each $f$ to the fiducial $f=1$ result. The final columns report the coefficient of determination for the source-fraction regression, including the pure basis elements.}
    \label{tab:oldagb_sensitivity}
    \begin{tabular}{c cccc cccc}
        \hline
        & \multicolumn{4}{c}{Gas} & \multicolumn{4}{c}{Mass-weighted stars} \\
        $f$ & median $|\Delta\xi|$ & max $|\Delta\xi|$ & Spearman $\rho$ & $R^2$ & median $|\Delta\xi|$ & max $|\Delta\xi|$ & Spearman $\rho$ & $R^2$ \\
        \hline
        0   & 0.078 & 0.181 & 0.926 & 0.70 & 0.028 & 0.132 & 0.908 & 0.77 \\
        0.5 & 0.028 & 0.082 & 0.993 & 0.81 & 0.007 & 0.023 & 0.993 & 0.71 \\
        1   & --    & --    & 1.000 & 0.84 & --    & --    & 1.000 & 0.73 \\
        2   & 0.041 & 0.135 & 0.979 & 0.88 & 0.010 & 0.028 & 0.977 & 0.77 \\
        \hline
    \end{tabular}
\end{table*}

\begin{figure*}
    \centering
    \includegraphics[width=1\textwidth]{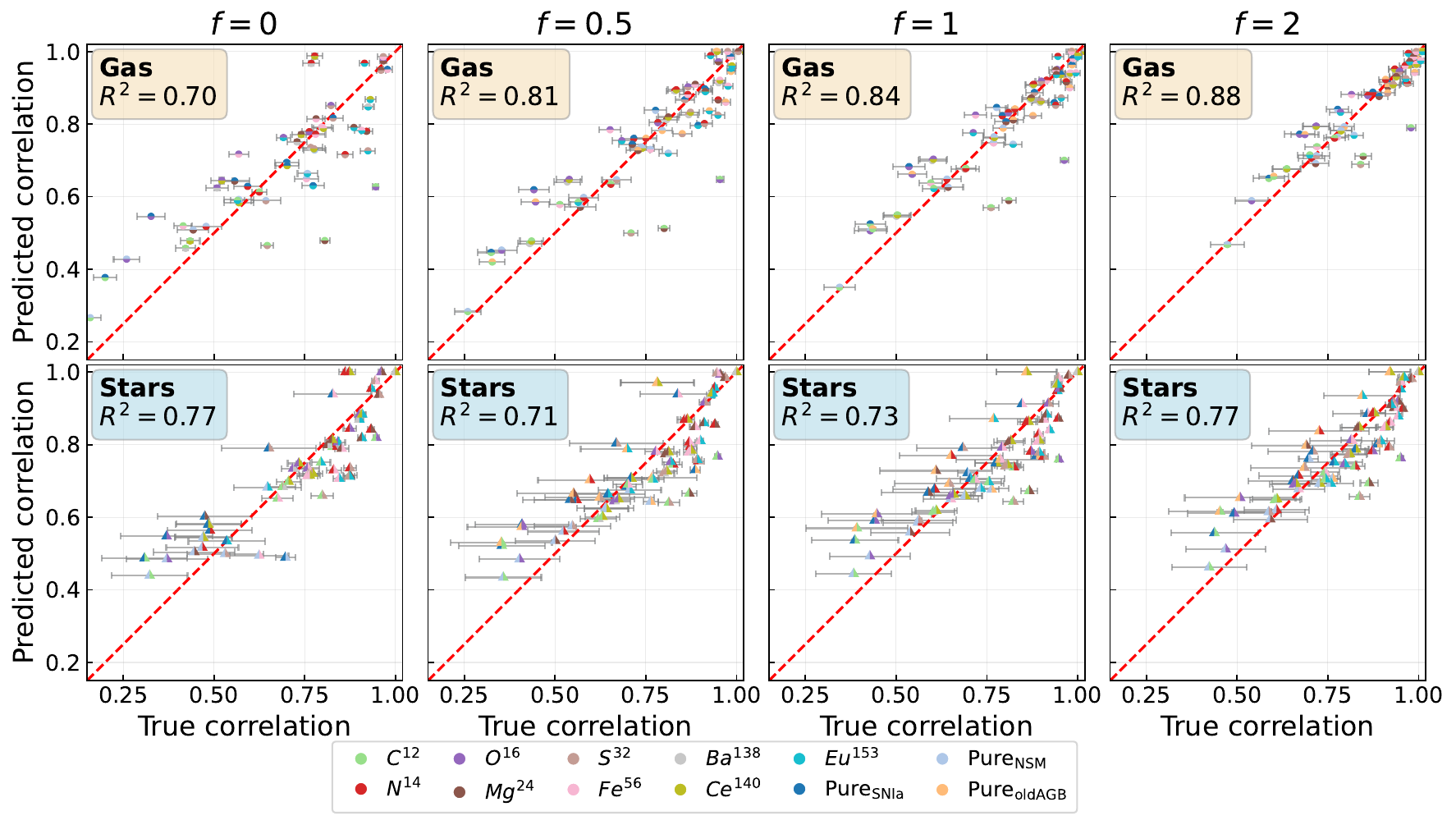}
    \caption{Regression predictions under the four old-star AGB yield multipliers. Columns show $f=0$, $0.5$, $1$, and $2$; rows show gas (half-circle markers) and mass-weighted stellar correlations (triangular markers). Marker colours, error bars, and the one-to-one line follow \autoref{fig:linear_regress}. The three pure elements are included for $f>0$; at $f=0$, $\mathrm{Pure}_{\mathrm{oldAGB}}$ and the inactive oldAGB regression term are omitted.}
    \label{fig:oldagb_sensitivity_regression}
\end{figure*}

The comparison in \autoref{tab:oldagb_sensitivity} separates quantitative sensitivity from robustness of the physical interpretation. Removing the old-star AGB contribution altogether gives the largest departure from the fiducial correlation values, particularly in the gas, but the element-pair rank ordering remains strongly correlated with the fiducial result ($\rho=0.926$ for gas and $0.908$ for stars). For the less extreme $f=0.5$ and $2$ cases, $\rho\geq0.977$ in both phases. The regression in \autoref{fig:oldagb_sensitivity_regression} retains substantial explanatory power throughout the tested range, with $R^2=0.70$--0.88 for gas and $0.71$--0.77 for stars. We therefore conclude that the precise numerical values of some gas-phase correlations are sensitive to the assumed old-star AGB return rate, but the main conclusions---that nucleosynthetic source similarity organises the correlation matrix, that same-source pairs correlate more strongly than mixed-source pairs, and that source-fraction differences predict much of the pair-to-pair variation---do not depend on the fiducial choice of this parameter. The mass-weighted stellar results are quantitatively more robust than the gas-phase results.

\section{Sources of observational discrepancies}
\label{app:obs_discrepancies}

In this appendix we provide a more detailed analysis of the inter-element correlation structure in the \citet{Casali2020} and \citet{GG2026} samples, and the reasons why the element-element correlations we extract from these two data sets might differ. Broadly speaking, these two data sets differ in that the former is a much smaller sample of stars selected to be both very nearby and very similar to the Sun, while the latter is a much broader sample, which offers far better statistics but at the price of larger statistical and systematic errors.  

Given the distinction between the two data sets, the first potential source of discrepancy to consider is abundance precision. Uncertainties are important because errors reduce the level of correlation. To test how much this effect contributes to the discrepancy between the two data sets, we degrade the \citeauthor{Casali2020} abundances (which have smaller quoted uncertainties) by adding Gaussian noise at the level of the median APOGEE uncertainties and recomputing the element-element correlations. We find that doing so moves the \citeauthor{Casali2020} correlations towards those of \citeauthor{GG2026}. Quantitatively, the mean absolute deviation (MAD) between the two correlation matrices is reduced from 0.139 to 0.089 when we add artificial noise to the \citeauthor{Casali2020} measurements. However, the fact that the residual remains 0.089 implies that precision alone cannot explain the offset.

A second potential source of systematic difference is the size of the window in [Fe/H] used to select the sample. The \citeauthor{Casali2020} sample is highly uniform and near-Solar, while the APOGEE sample used in \citeauthor{GG2026} is considerably broader. We can either compute correlations on the full sample, or down-select to pick out narrow [Fe/H] windows around Solar. This down-selection reduces the dynamical range in abundance space and thus can suppress correlations; for example, the MAD with the \citeauthor{Casali2020} sample is $0.14$ for the full \citeauthor{GG2026} sample, but this increases to $0.72$ and $0.56$ if we select stars within [Fe/H] windows of $\pm 0.05$ and $\pm 0.10$ dex, respectively. Thus differences in [Fe/H] coverage and intrinsic chemical diversity are likely an important contributor to differences between the two data sets.

A third potential origin for the disagreement between the two data sets is known temperature-dependent systematics, particularly for Ce in the infrared. These can be mitigated but not eliminated by restricting both samples to a narrow $T_\mathrm{eff}$ range. Doing so reduces the difference in cross correlations between Ce and other elements: $\Delta\xi_\text{Ce-Fe}$ is reduced from $-0.12$ to $-0.09$, $\Delta\xi_\text{Ce-Mg}$ from $-0.17$ to $-0.15$, and $\Delta\xi_\text{Ce-S}$ from $-0.15$ to $-0.14$, where we define $\Delta\xi = \xi_{\text{Ness}} - \xi_{\text{Casali}}$). This exercise confirms that such systematics are real, yet they are not the dominant driver of the overall discrepancy.

\begin{figure*}
    \centering
    \includegraphics[width=1\textwidth]{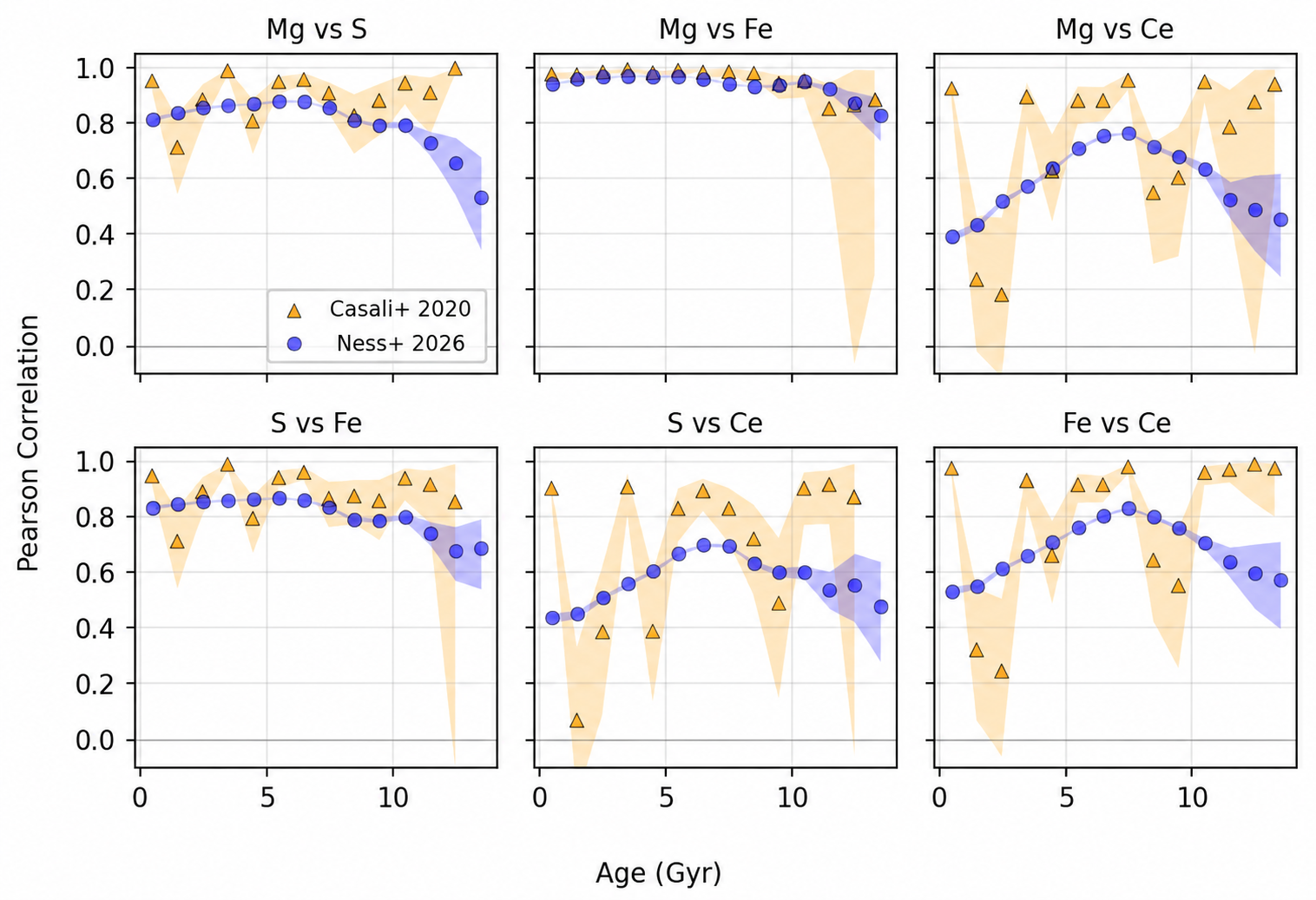}
    \caption{Cross-correlation as a function of stellar age for the six element pairs in common between the samples of \citet{Casali2020} and \citet{GG2026}. Points are computed in fixed 1 Gyr age bins (excluding bins with fewer than 5 samples). The shaded band around each sample trace shows the 95\% confidence interval of the Pearson correlation, estimated with the Fisher $z$ transform in each age bin and mapped back to correlation space. All panels share the same axes.}
    \label{fig:obs_analysis}
\end{figure*}

The fourth and most significant difference lies in the age estimates, which matter because abundance measurement errors and age estimates are often tightly coupled. \autoref{fig:obs_analysis} shows how the cross-correlations for six element pairs in common to the \citeauthor{Casali2020} and \citeauthor{GG2026} samples vary as a function of inferred stellar age. We see that the two data sets broadly agree at intermediate ages, but that they diverge substantially for stars older than $\sim 10$ Gyr. If we restrict the comparison to stars younger than 10 Gyr, the offsets between the two samples shrink considerably: $\Delta\xi_\text{Ce-Fe}$ changes from $-0.12$ to $-0.01$, $\Delta\xi_\text{Ce-Mg}$ from $-0.17$ to $-0.06$, $\Delta\xi_\text{Ce-S}$ from $-0.15$ to $-0.06$, $\Delta\xi_\text{Fe-S}$ from $-0.08$ to $-0.05$, and $\Delta\xi_\text{Mg-S}$ from $-0.10$ to $-0.04$; only $\Delta\xi_\text{Fe-Mg}$ remains unchanged at $-0.03$. This pattern indicates that much of the discrepancy between the two surveys is driven by the oldest stars and by the treatment of age. Moreover, the element pairs involving Ce in the \citeauthor{Casali2020} sample show particularly strong, non-monotonic fluctuations with age.
Perhaps most intriguingly, both samples independently exhibit a bump in cross-correlation for stars with ages $\sim 5$--$10$ Gyr, suggesting that the correlation structure is responding to features in the Galactic star formation history. While our simulations do not follow a full multi-Gyr evolution, the observed age trends in stellar abundances provide a powerful window onto the sequence of quasi-equilibrium states experienced by the Milky Way (and by analogy other nearby disc galaxies) over cosmic time.

In this sense, the higher correlations found in \citeauthor{Casali2020} compared to \citeauthor{GG2026} during the age range from $\sim 5$--10 Gyr may best be interpreted not as a failure of the methodology in either study, but as evidence that different surveys sample stellar populations with different effective chemical dimensionalities and measurement response functions, consistent with broader discussions of abundance precision and systematics  \citep[e.g.,][]{Jofre2019}.


\bsp	
\label{lastpage}
\end{document}